\documentclass[twocolumn,trackchange]{aastex62}

\newcommand{\HeII}{\ion{He}{2}\ensuremath{~\lambda}1640}
\usepackage{txfonts}
\graphicspath{{./}{figures/}}
\shorttitle{}
\shortauthors{Mondal et al.}

\begin{document}

\title{GNHeII~J1236+6215: A He~II~$\lambda$1640 emitting and potentially LyC leaking galaxy at $z$ = 2.9803 unveiled through JWST \& Keck observations}

\author[0000-0003-4531-0945]{Chayan Mondal}
\affiliation{Academia Sinica Institute of Astronomy and Astrophysics (ASIAA), No. 1, Section 4, Roosevelt Road, Taipei 106319, Taiwan}
\affiliation{Inter-University Centre for Astronomy and Astrophysics, Ganeshkhind, Post Bag 4, Pune 411007, India}

\author[0000-0002-8768-9298]{Kanak Saha}
\affiliation{Inter-University Centre for Astronomy and Astrophysics, Ganeshkhind, Post Bag 4, Pune 411007, India}

\author[0000-0002-2870-7716]{Anshuman Borgohain}
\affiliation{Inter-University Centre for Astronomy and Astrophysics, Ganeshkhind, Post Bag 4, Pune 411007, India}

\author[0000-0002-0648-1699]{Brent M. Smith}
\affiliation{School of Earth \& Space Exploration, Arizona State University, Tempe, AZ 85287-1404, USA}

\author[0000-0001-8156-6281]{Rogier A. Windhorst}
\affiliation{School of Earth \& Space Exploration, Arizona State University, Tempe, AZ 85287-1404, USA}

\author[0000-0001-9687-4973]{Naveen Reddy}
\affiliation{Department of Physics and Astronomy, University of California, Riverside, 900 University Avenue, Riverside, CA 92521, USA}

\author[0000-0002-3805-0789]{Chian-Chou Chen}
\affiliation{Academia Sinica Institute of Astronomy and Astrophysics (ASIAA), No. 1, Section 4, Roosevelt Road, Taipei 106319, Taiwan}

\author[0000-0002-7196-4822]{Keiichi Umetsu}
\affiliation{Academia Sinica Institute of Astronomy and Astrophysics (ASIAA), No. 1, Section 4, Roosevelt Road, Taipei 106319, Taiwan}

\author[0000-0003-1268-5230]{Rolf A.~Jansen}
\affiliation{School of Earth \& Space Exploration, Arizona State University, Tempe, AZ 85287-1404, USA}

\email{cmondal@asiaa.sinica.edu.tw, mondalchayan1991@gmail.com}

\begin{abstract}

He~II~$\lambda$1640 emission in galaxies indicates the presence of sources that produce extreme ionizing photons. Here, we report the discovery of a He~II~$\lambda$1640 emitting galaxy, GNHeII~J1236+6215, at $z=$ 2.9803 in the GOODS-north field. We use photometry in 17 wavebands from near-UV to infrared to characterize the galaxy SED and combine Keck LRIS and JWST NIRSpec spectra to identify 15 emission lines including He~II~$\lambda$1640. We infer that the He$^+$ ionization in the galaxy could be driven by small pockets of young Population III stars or low-metallicity Very Massive Stars (VMSs) rather than AGN or metal-rich Wolf-Rayet stars. The galaxy has a highly ionized ISM ([OIII]5007/[OII]3727 = 7.28$\pm$0.11, [SIII]/[SII] = 1.97$\pm$0.48 and detected Ly$\alpha$, H$\alpha$, H$\beta$, H$\gamma$ lines), little reddening by dust (E(B$-$V) = 0.04$\pm$0.12), low metallicity (12 + log(O/H) = 7.85$\pm$0.22), and high star formation rate (SFR$_{\rm SED}$ = 12.2$\pm$2.0 M$_{\odot}$ yr$^{-1}$). In addition to these ISM conditions, we also notice a significant [SII] deficiency ([SII]6718,6732/H$\alpha$ = 0.08$\pm$0.02, $\Delta$[SII] = $-$0.12) which may indicate the presence of density-bounded optically thin H~II regions that combined with the low dust extinction favor leaking of ionizing Lyman continuum (LyC) photons. Our best-fit SED model also infers a high nebular ionization (log~U = $-2.0$) and a low stellar mass M = 7.8$\pm3.1\times$10$^8$M$_{\odot}$. This discovery not only adds one important object to the known sample of high-redshift He~II emitters but also highlights a potential connection between He$^+$ ionization and favorable ISM conditions for the leakage of ionizing photons from galaxies.

\end{abstract}


\section{Introduction}
\label{s_intro}

The chemical enrichment of the universe kicked off with the evolution of the first generation of stars formed out of the primordial pristine gas. The formation of these metal-free stars (aka Population III stars, hereafter Pop III stars) follows a top-heavy stellar initial mass function (IMF) which results in their relatively higher mass and surface temperature \citep{schaerer2002,bromm2004,bromm2011}. Due to these unique properties, Pop~III stars are efficient contributors of energetic photons that can ionize hydrogen (E $> 13.6$eV i.e., $\lambda<912 \AA{}$) or even He$^+$ (E $>54.4$eV i.e., $\lambda<228 \AA{}$), and hence crucial for reionization studies \citep{tumlinson2000,schaerer2003,wise2014,lecroq2025}. The direct detection of ionizing photons from galaxies, especially beyond redshift 5.0, becomes prohibitive as the transmission of intergalactic medium falls below 10\% \citep{madau1995,inou2014}. However, considering the hard ionizing spectrum of Pop~III stars, detection of specific high ionization emission lines (eg., He~II~$\lambda$1640) from galaxies can be used to infer their presence instead.  

The combined detection of He~II~$\lambda$1640 and Ly$\alpha$ lines has been proposed as a signature of galaxies hosting Pop~III stellar populations \citep{tumlinson2001,schaerer2003}. Using stellar evolutionary models, \citet{schaerer2003} showed that He~II~$\lambda$1640 emission with an equivalent width $> 5 \AA{}$ is expected mostly from stars with metallicity Z $<10^{-7}$. However, both these recombination lines can also be generated by astrophysical sources other than Pop~III stars. \citet{yang2006} showed that the cooling of pristine metal-free gas, accreted to an over-density, can produce strong Ly$\alpha$ and He~II~$\lambda$1640 emission. The luminosity of these lines, estimated at $z\sim$ 2 - 3 by \citet{yang2006}, has a similar range to that produced by Pop~III star-forming regions of metallicity 0$<Z<10^{-5}$, thereby making it harder to distinguish between both cases. He~II~$\lambda$1640 and Ly$\alpha$ lines can also be produced in galaxies hosting massive Wolf-Rayet (WR) stars or an active galactic nucleus (AGN) \citep{leitherer1996,leitherer1999,crowther2007,cassata2013,themia2019,saxena2020}. However, unlike Pop~III stars, the spectra of WR galaxies or AGN also show bright C or N emission lines (eg., CIII]~$\lambda$1909, CIV~$\lambda$1549) in most cases \citep{leitherer1996,crowther2007,themia2019}, which help to differentiate the source of He$^+$ ionization. Besides, the He~II~$\lambda$1640 line produced by the stellar wind of WR stars or accretion disk of AGNs has relatively broader line widths (i.e., $>1000$; km~s$^{-1}$ \citealt{debreuck2000,cassata2013,saxena2020}) than what is expected in the case of Pop~III stars. However, \citet{grafener2015} showed that a narrower \HeII\ line (i.e., FWHM $\sim$ a few 100 km~s$^{-1}$), which is generally considered to be contributed by Pop~III stars \citep{sobral2015}, can also be powered by strong but slow WR-type stellar winds of low-metallicity very massive stars (VMS), typically more massive than 100 M$_{\odot}$ \citep{vink2015}. The presence of VMSs has been reported in nearby star-forming regions as well as in distant UV-bright galaxies \citep{crowther2010,smith2023,mestric2023,upadhyaya2024}. In addition to producing stronger \HeII\ line, VMSs of younger age can also significantly boost the UV luminosity and production efficiency of ionizing photons in galaxies \citep{schaerer2025}. The other notable astrophysical sources that can power He$^+$ ionization are stripped massive binary stars, winds driven by Supernova, and X-ray binaries in low metallicity environments \citep{allen2008,steidel2016,eldridge2017,schaere2019}. However, the choice of extreme IMF and binary stellar populations with extra sub-solar metallicity could not always reproduce the He~II line strength observed in both low and high-redshift galaxies \citep{stanway2019,themia2019}. Overall, 
galaxies hosting massive stars with increasingly low metallicity are more efficient in producing energetic photons that can power He$^+$ ionization \citep{schaerer2003}.

While the Pop~III stars are expected to be found in the epoch of cosmic dawn, several studies indicate that they could also form in lower redshift galaxies hosting pockets of pristine gas. Using numerical simulation, \citet{tornatore2007} showed that inefficient metal transport by outflow can still allow the formation of Pop~III stars down to redshift 2.5. Considering an optimistic scenario, \citet{liu2020} argued that the formation of Pop~III stars can continue even down to $z\sim0$ with a low but non-negligible rate. The discovery of two gas clouds with metallicity Z $<10^{-4}$ at $z\sim$3 emphasizes the possibility of such a detection at lower redshift \citep{fumagalli2011}. Though a confirmed identification of Pop~III stars is yet to happen, there have been several detections of galaxies showing He~II~$\lambda$1640 emission, which stands as one of the major signatures that Pop~III stars show. Regardless of the source, the detection of He~II~$\lambda$1640-emitter, both at local and distant universe, suggests the prevalence of extreme ionizing conditions in galaxies at different cosmic times. The deep spectroscopic data obtained using different instruments of the Very Large Telescope (VLT) have resulted in the detection of around 70 He~II~$\lambda$1640 emitters between redshifts 2 and 5 \citep{cassata2013,themia2019,saxena2020}. However, most of these galaxies also show other rest-UV emission lines, eg., CIV~$\lambda$1549, CIII]~$\lambda$1909, and OIII]~$\lambda$1661,1666, indicating the potential contribution of AGN or WR stars in a relatively metal-enriched interstellar medium (ISM), which argues against a pristine primordial environment. The detection of He~II~$\lambda$1640 emission is also reported in lensed metal-poor galaxies at higher redshift \citep{patricio2016,berg2018}. \citet{sobral2015} detected narrow He~II~$\lambda$1640 emission from a luminous Ly$\alpha$ emitter at the epoch of reionization. More recently, \citet{wang2024} reported the discovery of a He~II~$\lambda$1640 emitting galaxy at redshift 8.2, showing promising characteristics of Pop~III stars. Conversely, no convincing He~II~$\lambda$1640 emission was found in other searches of z$\gtrsim$4 galaxies \citep{nagao2005,nagao2008,shibuya2018}.

Similar extreme ionizing conditions have also been encountered in nearby galaxies. Hubble Space Telescope (HST)/Cosmic Origins Spectrograph (COS) UV spectra have revealed He~II~$\lambda$1640 lines in nearby star-forming galaxies \citep{senchyna2017,berg2019,liu2024}. The high ionization He~II~$\lambda$4686 line is identified in many H~II regions of local star-forming galaxies spanning a wider range of metallicity \citep{guseva2000,izotov2004,kehrig2011,shirazi2012,senchyna2019,kehrig2018,umeda2022,enders2023,mayya2023}. However, the He$^+$ ionization in these local sources is mostly attributed to massive WR stars, AGN, and X-ray binaries. 

Due to the uniqueness and relative scarcity of He~II~$\lambda$1640 emitters, detection of this emission line, especially from a metal-poor galaxy at higher redshift, is important to understand these sources better. In this study, we present the discovery of a He~II~$\lambda$1640 emitting galaxy, named as GNHeII~J1236+6215, at $z = 2.9803$ in the GOODS-north field using spectroscopic observations from the Keck and the James Webb Space Telescope (JWST). Unlike other known He~II emitters at similar redshift, we do not distinctly detect rest-UV/optical C or N lines but identify three more helium lines (i.e., He~II~$\lambda$8236, He~I~$\lambda$5875, He~I~$\lambda$10830) in this galaxy, indicating a less-enriched ISM. The derived He~II~$\lambda$1640 line properties, gas-phase metallicity, ionizing state of the ISM, dust content, star formation rate, and stellar mass of the galaxy not only favor the presence of extremely metal-poor populations but also resemble the conditions observed in galaxies that are known to leak ionizing radiation (eg., \citealt{barrow2020,wang2021,flury2022}). Our discovery strengthens the possibility of finding galaxies hosting Pop~III-like stellar populations even after the reionization epoch. We present the paper in the following order: In Section \ref{s_data} we discuss the utilized data and selection of the galaxy, Section \ref{s_detection} presents the methods of source detection and photometry, Section \ref{s_redshift_conf} discusses redshift confirmation of the galaxy, spectral energy distribution (SED) modeling in Section \ref{s_sed}, spectroscopic analysis in Section \ref{s_lines}, followed by discussion in Section \ref{s_discussion} and summary in Section \ref{s_summary}. Throughout the paper, we adopt a cosmology with $H_0 = 70$ km~s$^{-1}$ Mpc$^{-1}$, $\Omega_{\Lambda} = 0.7$, $\Omega_M = 0.3$\ and present all photometric magnitudes in the AB system \citep{oke1983}.

\begin{figure}
    \centering
    \includegraphics[width=3.5in]{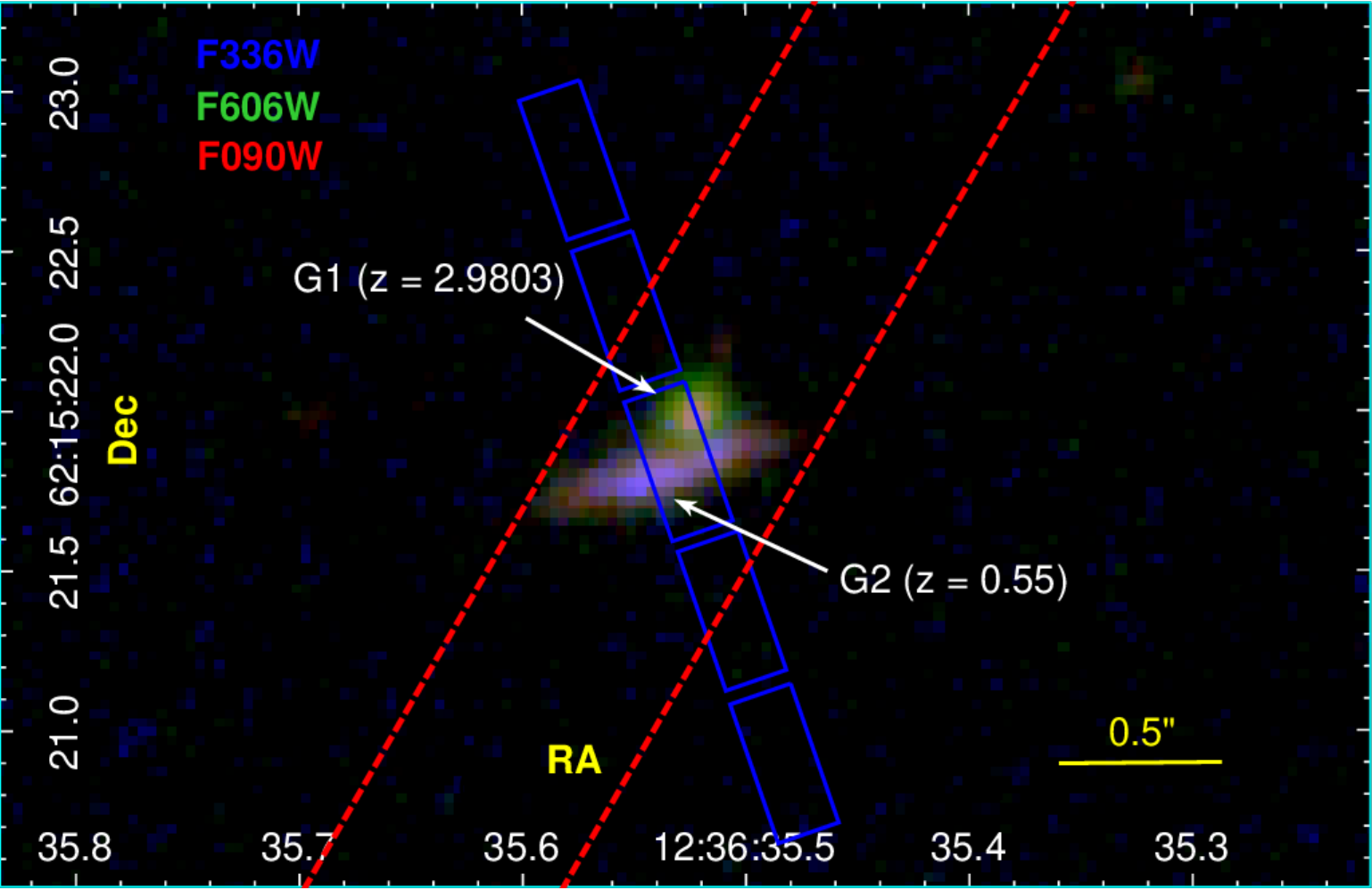}
    \caption{Color composite RGB image of the galaxies GNHeII~J1236+6215 (i.e., \HeII~emitter G1 at $z = 2.9803$) and G2 (i.e., foreground galaxy at $z = 0.55$). The emission in the HST F336W, HST F606W, and JWST F090W bands is displayed using the blue, green, and red colors, respectively. The red dashed lines show the orientation of the slit used in Keck MOSFIRE observations. Blue boxes show the alignment of the JWST NIRSpec MSA. The distinct color difference between the two galaxies signifies that G1 is not well-detected in the HST UV F336W band, while G2 shows emission in all the bands.}
    \label{fig:color_image}
\end{figure}

\section{Data and source selection}
\label{s_data}
We have utilized multi-band photometric and spectroscopic data to study a He~II~$\lambda$1640 emitting Lyman Break Galaxy (LBG) at $z=2.9803\pm0.001$ (Figure \ref{fig:color_image}) located in the GOODS-north deep field \citep{giavalisco2004}, selected from a catalog of LBGs at $z\sim3$ studied by \citet{steidel2003}. The galaxy has been observed by different telescopes from the X-ray to radio wavelengths \citep{alexander2003,capak2004,perez2013,kajisawa2011,ashby2013,dickinson2003,elbaz2011,magnelli2013,morrison2010,guidetti2017,rieke2023}. We initially identified this object as a Lyman Continuum (LyC) leaker, where the ionizing LyC photons (i.e., $\lambda<912 \AA{}$) were detected in the N242W band (bandpass $\sim$ 1700- 3000 $\AA$) image of the AstroSat UV Deep Field north (AUDFn; \citealt{Mondal2023a}) observed by the Ultra-Violet Imaging Telescope (UVIT) \citep{kumar2012}. However, the UVIT's point spread function (FWHM $\sim$ 1\farcs2) could not deblend a foreground galaxy from the LBG (discussed in Section \ref{s_detection}), which became clearer in the recently released JWST NIRCam images (FWHM$\sim$0\farcs033 (F090W)). Due to this foreground contamination, the ionizing photon escape measurements from this galaxy alone remain unreliable. In the remainder of this work, we highlight the He~II~$\lambda$1640 emission and the unique characteristics of this LBG at this redshift.

Following the photometric LBG classification, follow-up spectroscopy was obtained by \citet{steidel2003,reddy2006} with Keck optical (LRIS) and infrared (MOSFIRE) spectrographs. The Low-Resolution Imaging Spectrometer (LRIS; \citealt{oke1995}) of Keck is designed to operate in the optical window (wavelength range$\sim$ 0.31 - 0.95 $\mu$m; R $\sim$ 1000 at 5000~$\AA$), whereas the Multi-Object Spectrometer For Infra-Red Exploration (MOSFIRE; \citealt{mclean2012}) functions in the near-infrared band (wavelength range: 0.97  - 2.45 $\mu$m; multi-object slit spectroscopy with R $\sim$ 3500). For the galaxy GNHeII~J1236+6215, the spectroscopic redshift, derived using multiple identified emission lines, is reported as 2.9803 in the MOSFIRE Deep Evolution Field (MOSDEF) survey catalog (ID - GOODS-N/gn3-deep/23207) \citep{kriek2015}. The MOSDEF survey, compiled with the MOSFIRE spectrograph, was aimed to observe rest-frame optical spectra of galaxies between redshifts 1.37 and 3.80\,. We obtain the $H$ and $K_s$ band MOSDEF spectra of the galaxy from the \textit{MOSDEF data archive}\footnote{https://mosdef.astro.berkeley.edu/for-scientists/data-releases/}. \citet{reddy2006} reported the same redshift value, which they estimated using the Keck LRIS spectrum, for the targeted source (ID - BX1290). The LRIS spectrum, which picks up the He~II~$\lambda$1640 line in our study, is acquired through private communication. For more details on the LRIS spectroscopic observation, we refer to \citet{reddy2006}. 

We also use the recently released JWST NIRSpec spectra of the galaxy from the \textit{JWST Advanced Deep Extra-galactic Survey (JADES) data release}\footnote{https://archive.stsci.edu/hlsp/jades} \citep{eisenstein2023,rieke2023,deugenio2024}. The JADES NIRSpec data of the GOODS-north field \citep{deugenio2024} offers five different spectra of each targeted source: (1) one low resolution (R $\sim$ 30 -- 300) prism spectra covering the entire NIRSpec wavelength range from 0.6 -- 5.3 $\mu$m, (2) three medium resolution (R $\sim$ 500 -- 1500) spectra taken with the grating/filter combinations G140M/F070LP, G235M/F170LP, G395M/F290LP, and (3) a relatively higher resolution (R $\sim$ 2700) spectra acquired with the grating/filter G395H/F290LP. We utilize both 1D and 2D spectra taken with the prism and three medium-resolution gratings (the G395H/F290LP spectrum has not been released yet) of our galaxy (NIRSpec ID = 00024755, TIER = goods-n-mediumhst). We note that the NIRSpec observation of our source is not affected by the Micro Shutter Assembly (MSA) short circuit (\textsc{DR\_FLAG = False}) and hence the spectra are robust for detailed scientific analysis. The NIRSpec spectroscopic catalog released by \citet{deugenio2024} reported redshift of the source as 2.9806$\pm$0.0002, estimated from multiple emission lines detected with S/N $>$ 5.

For photometric analysis, we have utilized the HST WFC3 UVIS and ACS images in 6 wavebands (F275W, F336W, F435W, F606W, F775W, F850LP) and JWST NIRCam images taken with 11 different filters (F090W, F115W, F150W, F182M, F200W, F210M, F277W, F335M, F356W, F410M, F444W). In Figure \ref{fig:color_image}, we show an HST-JWST color composite image of the galaxy. All the retrieved images, provided in the JADES archive, have the same pixel scale of size $\sim$0\farcs03. We have not used the HST WFC3 IR images, available with a pixel size 0\farcs06, as the galaxy suffers severe blending with the foreground source in those. The HST images, with a typical 5$\sigma$ depth of $\sim$ 27 -- 28 mag (measured with 0\farcs4 diameter aperture), are retrieved from the \textit{Hubble Legacy Fields Data Release V2.5
}\footnote{https://archive.stsci.edu/hlsps/hlf/v2.5/goodsn/30mas/} \citep{oesch2018}. The JWST NIRCam images are obtained from the same JADES data archive \citep{rieke2023}. The typical exposure time of the NIRCam images ranges from $\ sim$8.5 ksec (F410M) to $\ sim$22.5 ksec (F115W) with median 5$\sigma$ depths reaching $\ sim$29.2 -- 29.9 mag (measured with 0\farcs3 diameter aperture). As all the acquired images are background-subtracted, we directly use them to perform photometry and measure the source flux in each band.

\section{Source Detection and Photometry}
\label{s_detection}
\begin{figure*}
    \centering
    \includegraphics[width=6.5in]{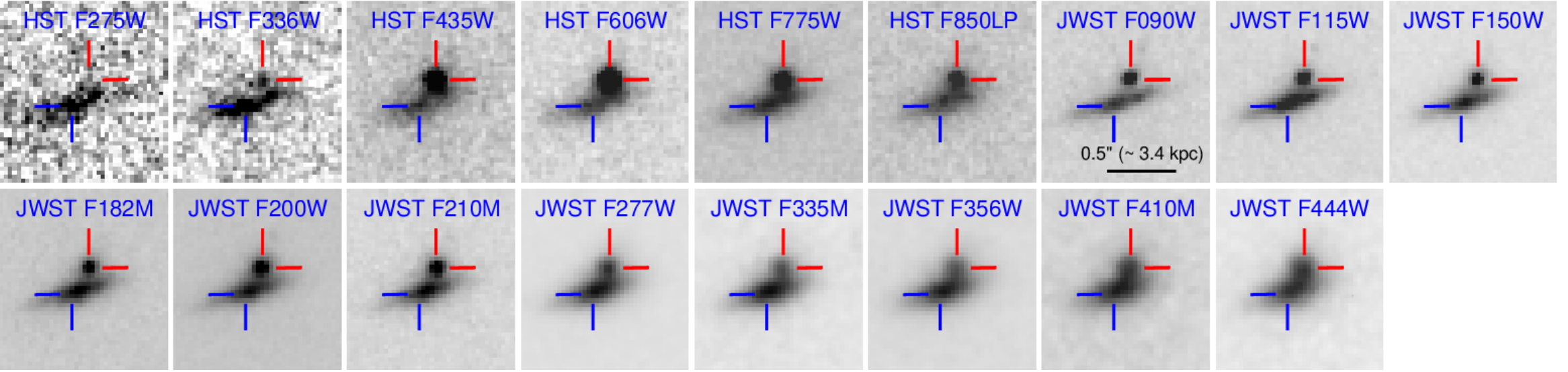}\\
    \includegraphics[width=6.5in]{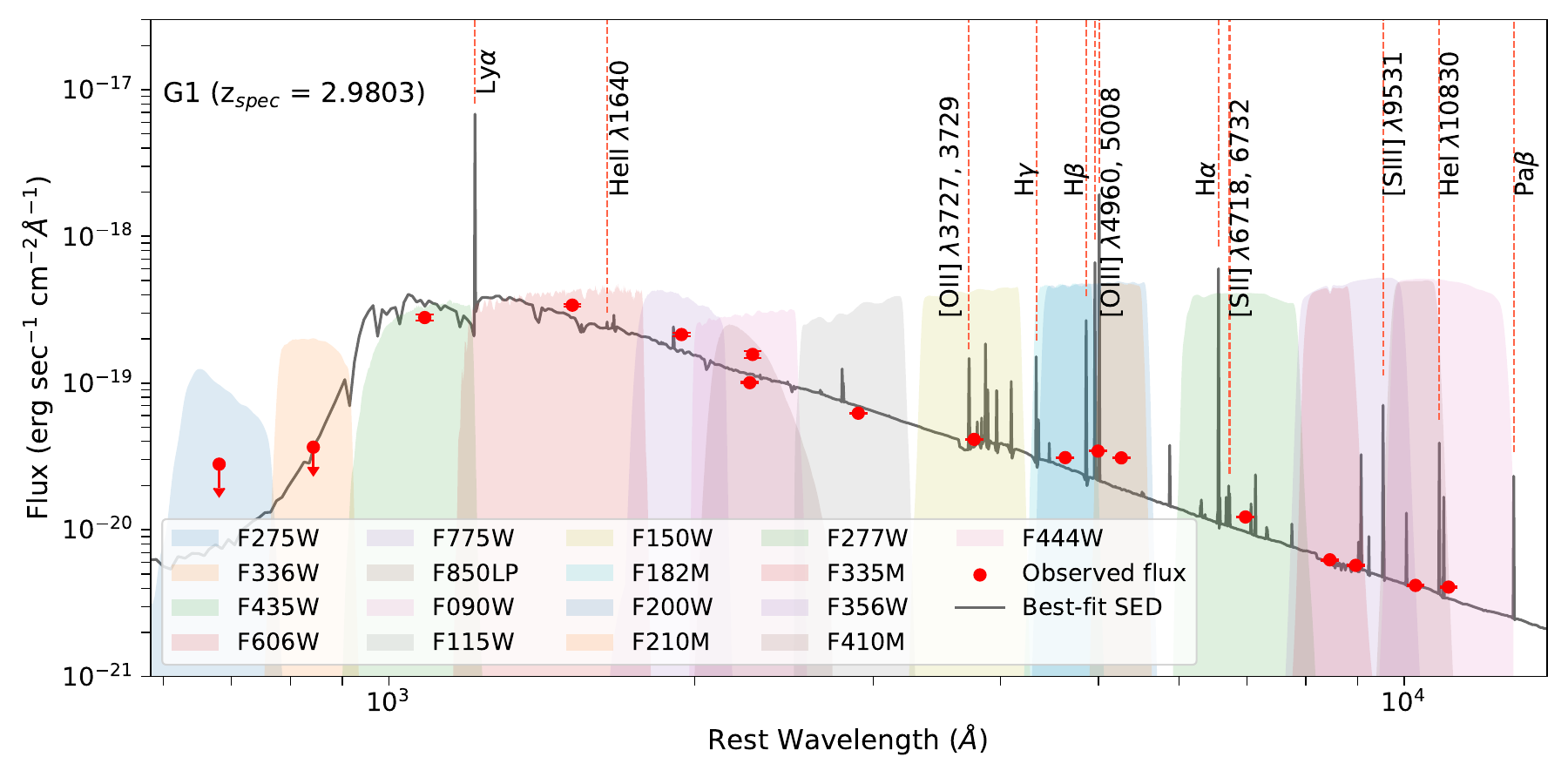}\\
    \includegraphics[width=6.5in]{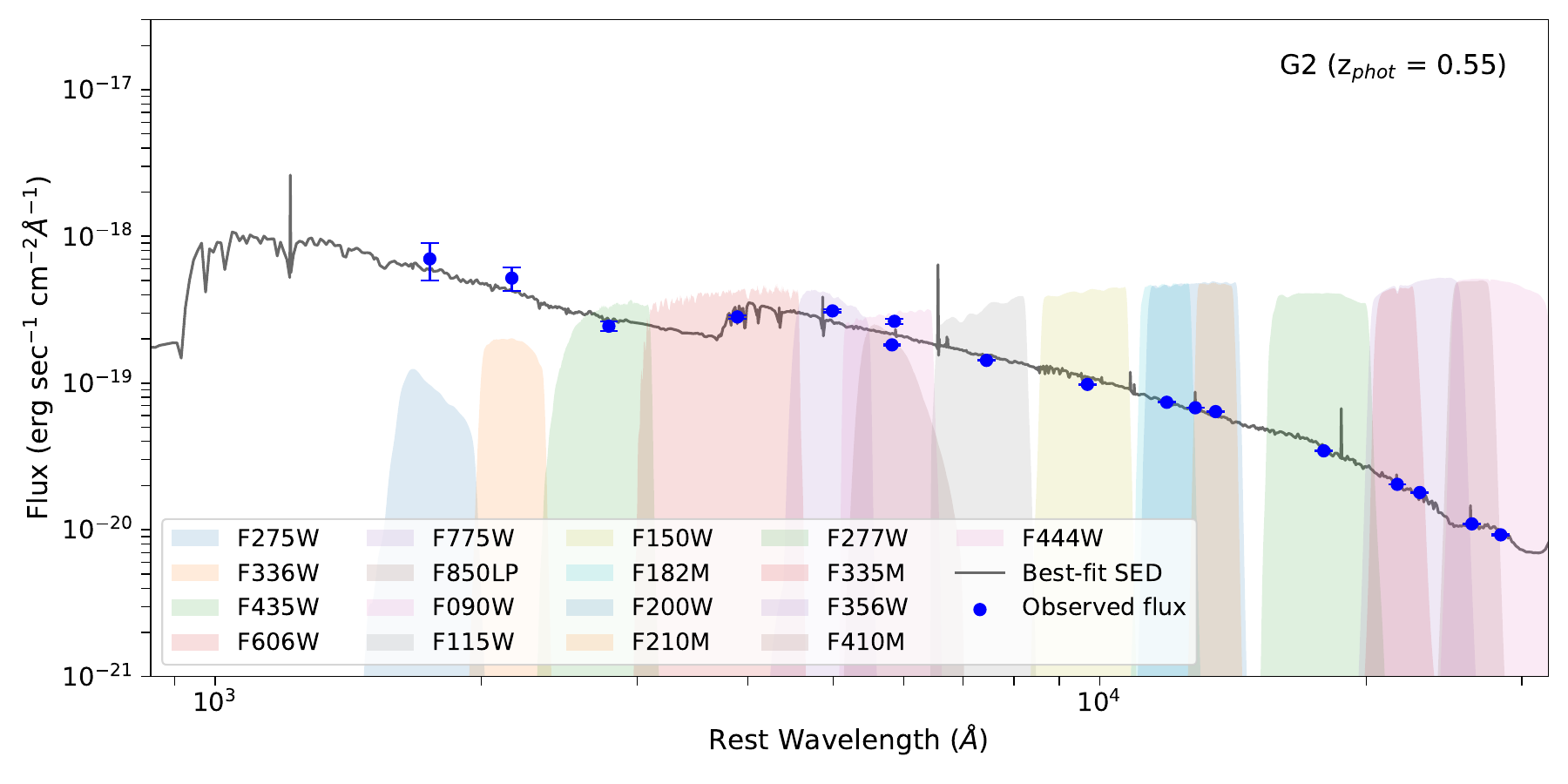} 
    \caption{(Top) 1\farcs15$\times$1\farcs25 image cut-outs of the galaxies G1 and G2 in 6 HST (F275W, F336W, F435W, F606W, F775W, F850LP) and 11 JWST (F090W, F115W, F150W, F182M, F200W, F210M, F277W, F335M, F356W, F410M, F444W) bands. The $z=2.9803$ He~II~$\lambda$1640 emitting galaxy GNHeII~J1236+6215 (G1) is marked in red, whereas the foreground galaxy G2 at $z=0.55$ is indicated by blue markers. A scale bar of 0\farcs5 ($\sim$3.4 kpc at the redshift of G1) is shown in the JWST F090W band image. The observed fluxes (red points (G1), blue points (G2)) and the CIGALE best-fit model SEDs (in greys) of the galaxy G1 and G2 are shown in the \textit{middle} and \textit{bottom} panels, respectively. Both SEDs are shown in the rest frame of each respective galaxy. The emission lines, identified in the observed spectra (discussed in Section \ref{s_lines}), are marked in the best-fit SED of G1 in the \textit{middle} panel.}
    \label{fig:image_sed}
\end{figure*}

The galaxy GNHeII~J1236+6215 has been listed in the 3D-HST photometric catalog (ID = 23207) by \citet{skelton2014}. This catalog was constructed using HST images with a pixel size 0\farcs06, in which the galaxy is blended with a nearby object (in projection) and identified there as a single source. Therefore, the photometric quantities listed in the 3D-HST catalog are biased. The source confusion reduces somewhat in the HST 0\farcs03 resolution images. However, the better PSF of the JWST NIRCam SW bands recovers both sources so that two individual objects are listed in the NIRCam photometric catalog released by \citet{rieke2023}. In the upper panel of Figure \ref{fig:image_sed}, we show the observed multi-band HST and JWST images of both sources. The point-like galaxy GNHeII~J1236+6215 (marked in red - hereafter we call it G1), which is not detected in the HST F275W band, confirming a $z\sim3$ LBG, is the source of our interest. The other source (marked in blue - hereafter we call it G2) is extended in nature and shows emission in all 17 photometric bands. Both sources are best resolved in the six JWST NIRCam SW bands, i.e., F090W, F115W, F150W, F182M, F200W, and F210M, which have PSF FWHM ranging between 0\farcs033 and 0\farcs071. In Table \ref{table:properties}, we provide some general information and important derived properties of both galaxies.

To estimate the flux of G1 and G2 in all the photometric bands, we first bring all the images to the same image resolution by performing PSF matching and convolution. We perform the photometry using T-PHOT v2.0 \citep{merlin2015,merlin2016}. A detailed description of the photometry is provided in Appendix \ref{s_tphot}. We list the derived flux values of both galaxies in Table \ref{table_photometry}.

\begin{table}
\caption{Properties of the He~II~$\lambda$1640 emitting galaxy GNHeII~J1236+6215 (G1) and the foreground galaxy (G2).}
\label{table:properties}
\begin{tabular}{p{3.3cm}p{2.2cm}p{2.2cm}}
\hline
Parameter & Value (G1) & Value (G2) \\\hline
\multicolumn{3}{c}{General parameters}\\\hline
RA [degree] & 189.1480105 & 189.1480556  \\
Dec [degree] & 62.2561051 & 62.2560601 \\
$z_{spec}$ & 2.9803$\pm$0.0010 & -\\
M$_{\rm 1500}$ & $-$22.09$\pm$0.02 & $-$16.76$\pm$0.29\tablenotemark{a}\\\hline
\multicolumn{3}{c}{From photometric analysis}\\\hline
 $z_{phot}$ & 2.897$^{+0.331}_{-0.402}$ & 0.55$^{+0.10}_{-0.08}$\\
 Stellar Mass [M$_{\odot}$] & (7.8$\pm$3.1)$\times$10$^8$  & (2.4$\pm$0.4)$\times$10$^8$\\
 $\beta_{obs}$ & $-2.18\pm0.06$ & $-1.73\pm0.09$\\
 SFR (10 Myr)[M$_{\odot}$ yr$^{-1}$] & 12.2$\pm$2.0 & 0.23$\pm$0.07\\\hline
\multicolumn{3}{c}{From spectroscopic analysis}\\\hline
SFR$_{UV}$ [M$_{\odot}$ yr$^{-1}$] & 9.8$\pm$0.1 & - \\
SFR$_{H\beta}$ [M$_{\odot}$ yr$^{-1}$] & 7.6$\pm$0.4 & -\\
SFR$_{H\alpha}$ [M$_{\odot}$ yr$^{-1}$] & 7.5$\pm$0.1 & - \\
SFR$_{Pa\beta}$ [M$_{\odot}$ yr$^{-1}$] & 6.40$\pm$0.03 & -\\
12 + log(O/H) & 7.85$\pm$0.22 & -\\
E(B$-$V) & 0.04$\pm$0.12 & \\
  f$_{[{\rm OIII}]}$/f$_{[{\rm OII}]}$ (O32) & 7.28$\pm$0.11 & -\\
  f$_{[{\rm SIII}]}$/f$_{[{\rm SII}]}$ ([SIII]/[SII])\tablenotemark{b} & 1.97$\pm$0.48 & -\\
  f$_{[{\rm SII}]}$/f$_{{\rm H{\alpha}}}$ & 0.08$\pm$0.02 & -\\
  f$_{esc}^{{\rm Ly\alpha}}$ & 0.16$\pm$0.03 & -\\\hline
\end{tabular}
\tablenotetext{a}{Estimated using observed F275W magnitude which corresponds to rest-frame wavelength of $\sim$1750~$\AA{}$ for G2.}
\tablenotetext{b}{[SIII]/[SII] is defined as [SIII]~$\lambda$9532,9069/[SII]~$\lambda$6718,6732}
\end{table}

\begin{table}
\caption{Details of the photometric bands and observed magnitudes of both galaxies in the HST and JWST bands}
\label{table_photometry}
\begin{tabular}{p{2.3cm}p{1.3cm}p{1.9cm}p{1.9cm}}
\hline
Filter & Effective & Magnitude & Magnitude\\
 & wavelength ($\AA$) & of G1 & of G2 \\\hline 

WFC3/F275W	&	2709	&	29.313$\pm$2.161	&	25.816$\pm$0.332\\
WFC3/F336W	&	3354	&	28.563$\pm$1.251	&	25.683$\pm$0.201\\
ACS/F435W	&	4318	&	25.845$\pm$0.057	&	25.993$\pm$0.079\\
ACS/F606W	&	6034	&	24.952$\pm$0.020	&	25.153$\pm$0.030\\
ACS/F775W	&	7729	&	24.886$\pm$0.029	&	24.482$\pm$0.024\\
ACS/F850LP	&	9082	&	24.860$\pm$0.062	&	24.295$\pm$0.045\\
NIRCam/F090W	&	9024	&	25.324$\pm$0.014	&	24.680$\pm$0.009\\
NIRCam/F115W	&	11541	&	25.305$\pm$0.011	&	24.405$\pm$0.006\\
NIRCam/F150W	&	15009	&	25.178$\pm$0.010	&	24.240$\pm$0.005\\
NIRCam/F182M	&	18452	&	25.038$\pm$0.011	&	24.095$\pm$0.006\\
NIRCam/F200W	&	19876	&	24.765$\pm$0.007	&	24.026$\pm$0.004\\
NIRCam/F210W	&	20956	&	24.765$\pm$0.010	&	23.977$\pm$0.006\\
NIRCam/F277W	&	27769	&	25.159$\pm$0.009	&	24.032$\pm$0.004\\
NIRCam/F335M	&	33624	&	25.474$\pm$0.014	&	24.187$\pm$0.005\\
NIRCam/F356W	&	35654	&	25.441$\pm$0.011	&	24.199$\pm$0.004\\
NIRCam/F410M	&	40836	&	25.487$\pm$0.016	&	24.439$\pm$0.008\\
NIRCam/F444W	&	44018	&	25.352$\pm$0.018	&	24.463$\pm$0.010\\

\hline
\end{tabular}

\textbf{Note.} Table columns: (1) name of the photometric filter; (2) effective wavelength of the filter in \AA~; (3) magnitude of the galaxy G1 in AB system; (4) magnitude of the galaxy G2 in AB system.
\end{table}

\begin{figure}
    \centering
    \includegraphics[width=3.5in]{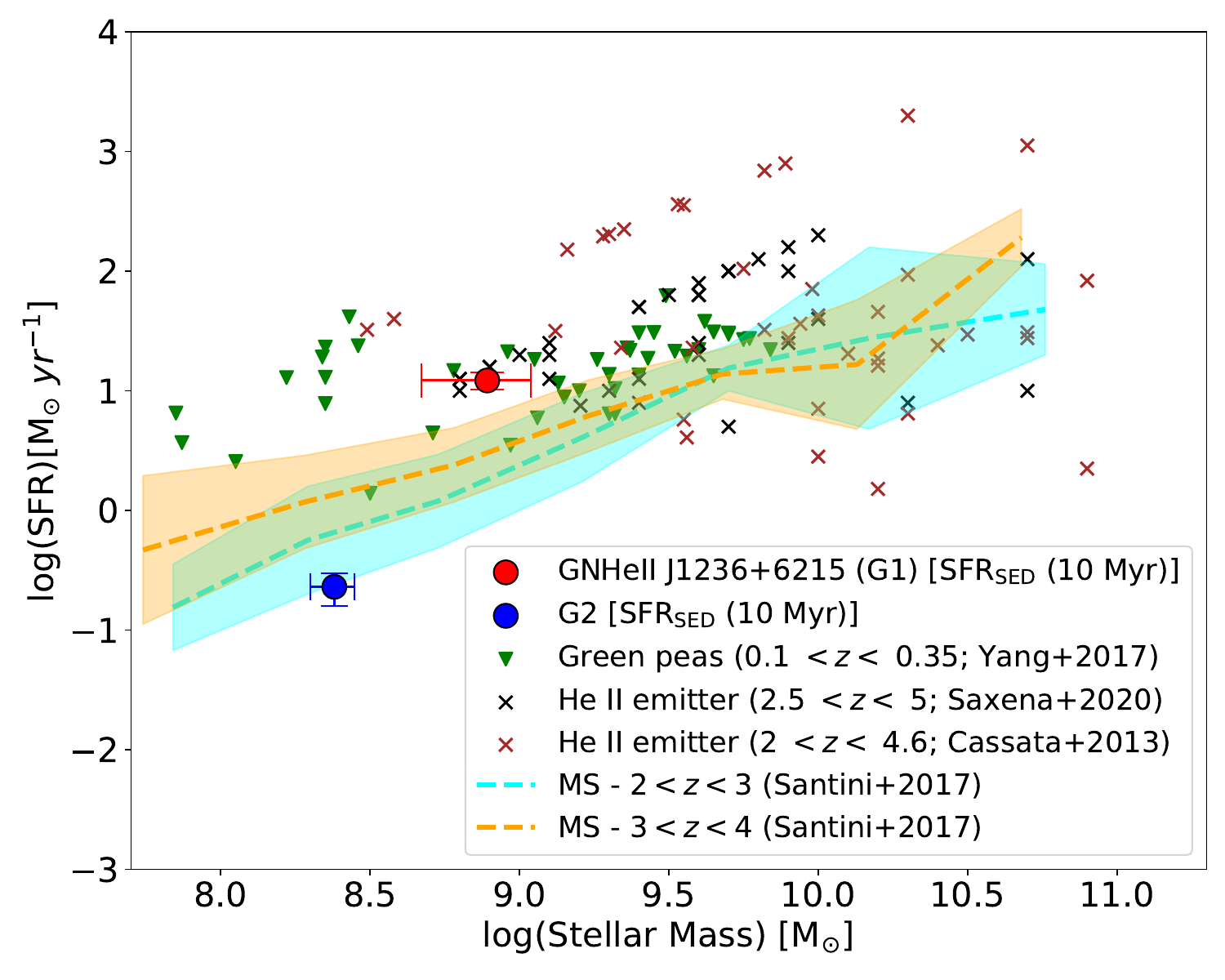}    
    \caption{The derived SFR and stellar mass of G1 and G2 are shown by the red and blue markers, respectively. The main sequence (MS) SFR-stellar mass relations derived for galaxies at redshift 2$<z<$3 and 3$<z<$4 are shown by the cyan and orange dashed lines from \citet{santini2017}. The shaded region around each respective line marks the 1$\sigma$ scatter in the SFR. The SFR and stellar mass of other known He~II emitters are shown in black (\citealt{saxena2020}) and brown (\citealt{cassata2013}) crosses. The green triangles represent Green pea galaxies at redshift 0.1 $< z <$ 0.35 \citep{yang2017}.}
    \label{fig:sfr_mass}
\end{figure}

\section{Redshift confirmation}
\label{s_redshift_conf}

We use our multi-band flux measurement, after correcting for Galactic extinction (as described in section \ref{s_sed}), to derive photometric redshift for both galaxies using \textit{EAZY} \citep{brammer2008}. The SED of G1 in the observed frame shows a break in the HST UVIS bands, whereas G2 has increasing flux values towards the shorter wavelength (Figure \ref{fig:image_sed}). The best-fit template of \textit{EAZY} predicts a photometric redshift of 0.55$^{+0.10}_{-0.08}$ for G2, whereas the photometric redshift of G1 ($z_{phot}$ = 2.897$^{+0.331}_{-0.402}$) agrees closely with the spectroscopic redshift derived from the Keck and JWST spectra \citep{steidel2003,reddy2006,kriek2015,deugenio2024}. However, the slit-width used in Keck MOSFIRE observations entirely encloses both G1 and G2 inside, whereas the orientation of the NIRSpec MSA shutters covers both galaxies partially (Figure \ref{fig:color_image}). Therefore, all the detected emission lines that closely support the photometric redshift of G1 must be checked for an alternative solution supporting the photometric redshift of G2. To do that, we consider a list of well-known emission lines from far-UV to near-IR wavelength range and try finding another redshift solution for which we could identify at least two from the list of all 15 detected emission lines (discussed in Section \ref{s_lines}). Following a rigorous check, we failed to find any solution other than $z = 2.9803$ to validate the detected lines. We specifically try to find a low redshift solution for the two lines (i.e., Ly$\alpha~\lambda$1215 and He~II~$\lambda$1640 as per $z = 2.9803$) detected in LRIS optical spectra. We got a negative result for this as well. This exercise confirms that all the 15 emission lines detected in the spectra obtained from different telescopes are produced by the galaxy G1. Furthermore, utilizing the JWST 2D spectra, we investigate the nature of the continuum and lines spatially along the length of the MSA and find convincing proof for G1 at $z=2.9803$ to be the source of the detected emission lines. We discuss this in detail in Appendix \ref{s_jwst_2d}.

We further assume that the photometric redshift of G2 is not robust, and it might actually be located at $z$ = 2.9803 along with G1 and contributing to the He~II~$\lambda$1640 line together. To investigate this scenario, we estimate the surface brightness of both galaxies in the HST F606W band, which probes the rest-frame FUV continuum at $z=2.9803$. The surface brightness of galaxy G2 turns out to be around 1.2 mag fainter compared to that of galaxy G1. This could also indicate that the detected UV lines are less likely to have originated from the galaxy G2. Therefore, we conclude that the galaxy G1 is located at $z = 2.9803$ (i.e., the value reported in both Keck and JWST catalogs), whereas G2 is a foreground galaxy at $z$ = 0.55 (according to our Phot-z solution).

\begin{table}
\caption{The CIGALE input parameters used for the SED modeling of the He~II~$\lambda$1640 emitting galaxy GNHeII~J1236+6215. The values corresponding to the best-fit SED are shown in bold font.}
\label{table_cigale}

\begin{tabular}{p{3.0cm}p{5.0cm}}
\hline
Parameters & Values\\\hline
\multicolumn{2}{c}{SFH - sfh2exp}\\\hline

Age (Myr) & 200, \textbf{500}, 1000, 1500, 2000\\
$\tau_{main}$ (Myr) & 100, 200, \textbf{500}, 1000, 2000\\
Burst$_{age}$ (Myr) & 1, 3, \textbf{5}, 10, 50, 100\\
$\tau_{burst}$ (Myr) & \textbf{5}, 10, 50, 100\\
$f_{burst}$ & 0.04, 0.06, 0.08, \textbf{0.1}, 0.2\\\hline

\multicolumn{2}{c}{SSP model - BC03 \citep{bruzual2003}}\\\hline

Stellar IMF & Chabrier \citep{chabrier2003}\\
Metallicity & 0.004, \textbf{0.008}\\\hline

\multicolumn{2}{c}{Nebular}\\\hline
logU & $-$4.0, $-$3.0, \textbf{$-$2.0}\\
Z$_{Gas}$ & 0.001, \textbf{0.003}, 0.004, 0.008\\
n$_e$ & 100\\
f$_{\rm esc}$ & 0.1, 0.2, 0.3, \textbf{0.4}, 0.5, 0.7, 0.9\\\hline

\multicolumn{2}{c}{Dust attenuation law}\\\hline
Dust law & Modified Calzetti 2000 \citep{calzetti2000}\\
E(B$-$V)$_{line}$ & 0.02, 0.07, \textbf{0.15}\\
E(B$-$V) factor & 0.44\\
Power law slope ($\delta$) & \textbf{-0.5}, 0.0\\\hline
\end{tabular}
\end{table}

\begin{figure*}
    \centering
    \includegraphics[width=7in]{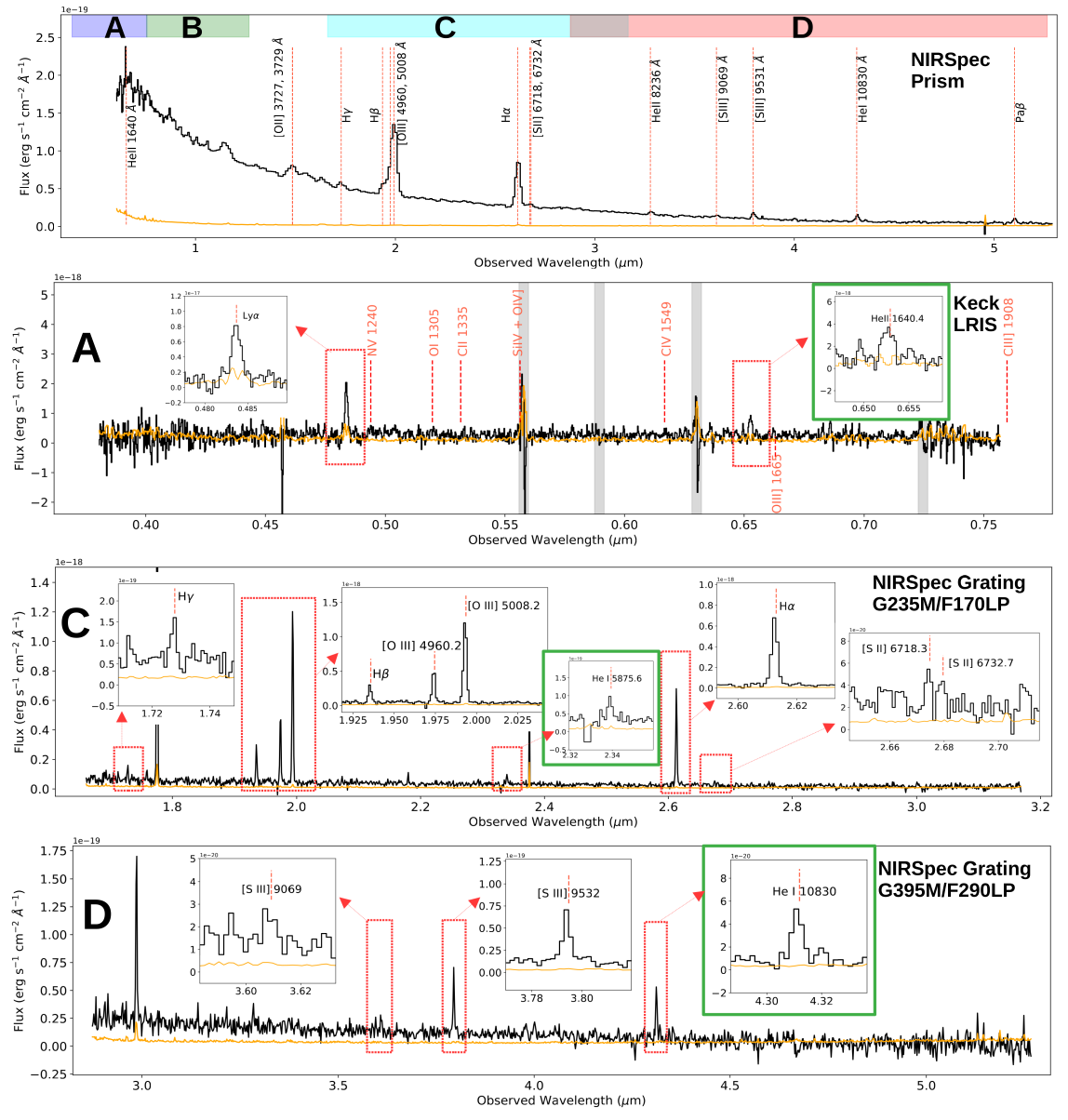}
    \caption{The JWST and Keck spectra that contain all the identified emission lines. In all the panels, the observed spectral fluxes are shown in black, while the errors are displayed in orange. The rectangular patches shown on the top panel using blue (A), green (B), cyan (C), and red (D) colors represent the wavelength coverage of Keck LRIS, JWST gratings G140M/F070LP, G235M/F170LP, and G395M/F290LP spectra, respectively, which are shown in the lower panels (except the G140M/F070LP spectrum which is shown in Figure \ref{fig:jwst_f070lp}). All the emission lines (except Ly$\alpha$, which falls outside the prism's wavelength range), identified in the LRIS and JWST grating spectra, are marked in vertical red dashed lines (top panel). Three additional lines, i.e., [OII]~$\lambda$3727/3729, He~II~$\lambda$8236, and Pa$\beta$, which are either not covered or well-identified in the JWST MSA grating spectra, are detected in the JWST prism spectrum. In the remaining three panels (2nd - LRIS (A), 3rd - G235M/F170LP (C), 4th - G395M/F290LP (D)), we show the observed spectra with the detected emission lines marked by red boxes. The spectral regions around each line are shown in the inset. The wavelength corresponding to the line peak is marked with a vertical red dashed line. The grey vertical patches in the LRIS spectra (block A) signify the spectral region of some optical sky lines. The insets indicating three identified helium lines are highlighted by green boxes.}
    \label{fig:all_spectra}
\end{figure*}

\section{SED modeling}
\label{s_sed}
Considering the redshift of G1 ($z_{spec} = 2.9803$) and G2 ($z_{phot} = 0.55$), we perform SED modeling of both galaxies using \textsc{CIGALE} \citep{boquien2019}. We utilize photometry in 17 wavebands for both galaxies after correcting for Galactic extinction assuming Fitzpatrick law \citep{fitzpatrick1999} and adopting V-band line of sight extinction (A$_V$) from the \textit{NASA/IPAC Infrared Science Archive} \footnote{https://irsa.ipac.caltech.edu/applications/DUST/} for the Schlegel Galactic extinction map \citep{schlegel1998}. We construct two optimal sets of input parameters corresponding to the redshift of each galaxy. For the galaxy G1, we constrain values of some input parameters from the emission line measurements. We derive internal reddening E(B$-$V) = 0.04$\pm$0.12 and metallicity Z $\sim$ 0.003$\pm$0.002 utilizing the value of Balmer decrement and S23 index, respectively (discussed in Section \ref{s_properties}). Considering these values and the associated error, we adopt a narrower range (from the allowed CIGALE values) for both E(B$-$V) and metallicity (as listed in Table \ref{table_cigale}) for the SED modeling of G1. We use BC03 stellar population models \citep{bruzual2003} with a Chabrier IMF \citep{chabrier2003} and consider an exponential star formation history added with a recent burst of varying strengths. All these parameters including the dust laws are specified in Table \ref{table_cigale}. The derived best-fit model SED (having reduced $\chi^2$ = 1.3) is shown in Figure \ref{fig:image_sed} along with the observed flux values. We note here that the estimated fluxes of G1 in both HST UVIS bands fall below the 5$\sigma$ detection limit and hence we show them as upper limits. The best-fit model confirms a recent starburst (burst age = 5 Myr) in the galaxy G1 with a star formation rate of 12.2$\pm$2.0 M$_{\odot}$ yr$^{-1}$ averaged over the last 10 Myr. The model-derived stellar mass of the galaxy is found as 7.8$\pm3.1\times$10$^8$ M$_{\odot}$, which confirms G1 to be a low-mass star-forming galaxy. The observed UV continuum slope estimated from the best-fit SED is $-2.18\pm0.06$, which indicates the presence of young stellar populations and lower dust in the galaxy.

\begin{table*}
\caption{Derived parameters of the identified emission lines}
\label{table_lines}
\begin{tabular}{p{2cm}p{2.cm}p{3.0cm}p{1.5cm}p{2cm}p{1.5cm}p{3.5cm}}
\hline

Emission & Wavelength & Observed flux & Observed & FWHM & Rest-frame & Instrument\\
Line & (\AA{}) & $\times 10^{-18}$ (erg~s$^{-1}$cm$^{-2}$) & FWHM (\AA{}) & (km~s$^{-1}$) & EW (\AA{}) & \\
(1) & (2) & (3) & (4) & (5) & (6) & (7)\\\hline
Ly$\alpha$ & 1215.48 & 23.0$\pm$5.4 & 12.2$\pm$1.4 & 758$\pm$90 (690) & 19.2 & Keck/LRIS\\
He~II~$\lambda$1640 & 1640.00 & 8.8$\pm$1.8 & 12.5$\pm$4.2 & 573$\pm$191 (526) & 8.3 & Keck/LRIS\\

[OIII]~$\lambda$4959 & 4960.1 & 10.7$\pm$2.3 & 6.5$\pm$1.9 & 99$\pm$29 (52) & 49.8 & Keck/MOSFIRE\\

[OIII]~$\lambda$5007 & 5008.3 & 59.6$\pm$2.2 & 11.2$\pm$0.6 & 169$\pm$10 (146) & 248.8 & Keck/MOSFIRE\\\hline

H$\gamma$ & 4341.5 & 1.79$\pm$0.35 & 17.1$\pm$23.4 & 296$\pm$407\tablenotemark{*} & 7.0 & NIRSpec G235M/F170LP\\
H$\beta$ & 4863.4 & 5.44$\pm$0.37 & 20.7$\pm$8.9 & 320$\pm$137\tablenotemark{*} & 26.6 & NIRSpec G235M/F170LP\\

[OIII]~$\lambda$4959 & 4960.8 & 10.78$\pm$0.37 & 20.3$\pm$5.0 & 308$\pm$75\tablenotemark{*} & 55.1 & NIRSpec G235M/F170LP\\

[OIII]~$\lambda$5007 & 5008.6 & 29.86$\pm$0.52 & 22.9$\pm$1.8 & 344$\pm$26\tablenotemark{*} & 155.9 & NIRSpec G235M/F170LP\\

He~I~$\lambda$5875 & 5877.5 & 1.60$\pm$0.32 & 23.4$\pm$37.0 & 300$\pm$475\tablenotemark{*} & 12.2 & NIRSpec G235M/F170LP\\
H$\alpha$ & 6564.7 & 16.46$\pm$0.32 & 23.4$\pm$3.6 & 268$\pm$41\tablenotemark{*} & 166.5 & NIRSpec G235M/F170LP\\

[SII]~$\lambda$6718 & 6720.0 & 0.75$\pm$0.19 & 20.1$\pm$62.4 & 225$\pm$700\tablenotemark{*} & 8.0 & NIRSpec G235M/F170LP\\

[SII]~$\lambda$6732 & 6732.3 & 0.57$\pm$0.24 & 23.5$\pm$110.0 & 263$\pm$1241\tablenotemark{*} & 6.2 & NIRSpec G235M/F170LP\\

[SIII]~$\lambda$9069 & 9068.0 & 0.61$\pm$0.29 & 46.0$\pm$23.4 & 382$\pm$195 & 11.1 & NIRSpec G395M/F290LP\\

[SIII]~$\lambda$9532 & 9533.8 & 2.10$\pm$0.20 & 34.9$\pm$8.9 & 276$\pm$70\tablenotemark{*} & 44.8 & NIRSpec G395M/F290LP\\
He~I~$\lambda$10830 & 10833.4 & 2.20$\pm$0.31 & 44.7$\pm$5.0 & 311$\pm$34 & 82.1 & NIRSpec G395M/F290LP\\\hline

[OII]~$\lambda$3727/29 & 3736$\pm$8 & 3.88$\pm$0.01 & - & - & 13.9 & NIRSpec prism\\
Pa$\beta$ & 12827$\pm$3 & 1.07$\pm$0.01 & - & - & 53.4 & NIRSpec prism\\

\hline
\end{tabular}
\tablenotetext{*}{The FWHM values are smaller than the limit of instrumental spectral resolution at that wavelength}

\textbf{Note.} Table columns: (1) name of the emission line; (2) rest-frame central wavelength of the line in \AA{} as derived from the fitting; (3) line flux in erg~sec$^{-1}$~cm$^{-2}$; (4) observed FWHM including the fitting error in $\AA{}$ as estimated from the fitted gaussian profile; (5) line FWHM in km~s$^{-1}$ - the values in the parenthesis (if any) represents intrinsic FWHM; (6) rest-frame equivalent width of the line in \AA~; (7) The instrument used to obtain the corresponding spectrum. 

\end{table*}

We follow the same method to perform SED modeling of the foreground galaxy G2 using CIGALE. Considering its relatively lower redshift, we adjust the stellar population age and metallicity values in the CIGALE input parameter list. We consider E(B$-$V) values between 0.3 and 0.5 as found in a sample of UV-selected galaxies at $\sim$0.5 by \citet{mondal2023} in the GOODS-N field and choose a low value of ionization parameter (i.e., log~U = $-4.0$) as we don't identify emission lines from the galaxy G2. The best-fit model SED (reduced $\chi^2$ = 0.9) of G2 is shown in the bottom panel of Figure \ref{fig:image_sed} along with the observed photometry. The SED modeling infers a star formation rate (SFR) of 0.23$\pm$0.07 M$_{\odot}$ yr$^{-1}$ and the galaxy stellar mass as 2.4$\pm0.4\times$10$^8$ M$_{\odot}$. We show the derived SFR and stellar mass of G1 and G2 in Figure \ref{fig:sfr_mass}, which signifies G1 to be a star-forming galaxy located slightly above the main-sequence at $z\sim3$, whereas G2 appears as a relatively less active low-mass galaxy. Moreover, the derived lower value of G2's stellar mass makes it an outlier with respect to the galaxies at $z\sim0.55$ listed in the SDSS+WISE catalog by \citet{chang2015}.

\section{Spectroscopic Analysis}
\label{s_lines}

\subsection{Identified emission lines}
\label{s_identified_lines}
In the LRIS 1D spectrum (Figure \ref{fig:all_spectra}), we identify two rest-frame FUV emission lines, Ly$\alpha~\lambda$1215 and He~II~$\lambda$1640, with SNR $>$ 4. The other high-ionization FUV lines, eg., N~V~$\lambda$1240, C~II~$\lambda$1335, Si~IV~$\lambda$1402, C~IV~$\lambda$1549 (marked in the same figure), which are a signature of stellar wind from massive O-type as well as WR stars \citep{leitherer1996,crowther2007}, are not distinctly identified in the LRIS spectrum. The estimated 1$\sigma$ upper limit of these lines ranges between $\sim$1.0--1.4$\times 10^{-18}$ erg~s$^{-1}$cm$^{-2}$. We note here that the spectral window of the Si~IV~$\lambda$1402 overlaps with an optical sky line; hence, we could not confirm its detection by the LRIS instrument, whereas the C~III]~$\lambda$1908 line falls outside the wavelength coverage of the spectrum. The JWST NIRSpec prism spectrum covers a part of the LRIS spectrum including the He~II~$\lambda$1640 line (Figure \ref{fig:all_spectra}). However, due to the poor spectral resolution and higher noise at the bluer end of the JWST prism spectrum, we could not confirm a clean detection of the He~II~$\lambda$1640 and the other UV lines (Figure \ref{fig:all_spectra}). Considering the noise from the NIRSpec prism spectrum, we derive a 1$\sigma$ upper limit for the \HeII~line as $\sim1.6\times10^{-18}$ erg~s$^{-1}$cm$^{-2}$.

We identify 8 rest-frame optical emission lines (i.e., H$\gamma$, H$\beta$, [OIII]~$\lambda$4959, 5007, He~I$\lambda$5875, H$\alpha$, [SII]~$\lambda$6718, 6732) in the NIRSpec G235M/F170LP grating 1D spectrum and 3 more lines (i.e., [SIII]~$\lambda$9069, [SIII]~$\lambda$9532 and He~I~$\lambda$10830) in the G395M/F290LP 1D spectrum (Figure \ref{fig:all_spectra}). We do not detect any prominent lines in the G140M/F070LP grating spectrum (Figure \ref{fig:jwst_f070lp}) which covers a wavelength range $\ sim$1900 -- 3190 $\AA{}$ in the rest-frame of G1. In the low-resolution NIRSpec prism spectrum, we identify [OII]~$\lambda$3727, 3729 which is not covered within the wavelength range sampled by the NIRSpec gratings (Figure \ref{fig:all_spectra}). We also mark two additional lines, He~II~$\lambda$8236 and Pa$\beta$, in the prism spectrum, which are not well-detected in the G395M/F290LP grating spectrum.

\subsection{Line properties}
\label{s_line_properties}

We use the Astropy \textit{specutils} package for fitting the continuum to all the observed spectra and deriving line fluxes from the respective continuum-subtracted spectra. Each detected emission line is fitted with a symmetric Gaussian function using the \textit{lmfit} package. The integrated flux within the best-fit Gaussian profile provides the measure of the line flux. The line FWHM is also estimated using the best-fit model. We estimate the line equivalent width (EW) using the fitted continuum, which we sample from the same spectral region used for deriving respective line fluxes. We note here that the fitted continuum under all the identified lines is well above the 1$\sigma$ error of flux values. As G2 is also expected to contribute to the observed continuum flux of the 1D spectra, the actual values of line flux and EW, originated in G1, could be slightly biased than what we have derived using the observed continuum. The error in line flux is calculated from each corresponding error spectrum utilizing a spectral window of width 2$\times$ FWHM under each respective line. For the LRIS spectrum, with no archival error measurement, we estimate the error in fluxes from the standard deviation of five adjacent flux values at each wavelength. The estimated properties of all detected lines are listed in Table \ref{table_lines}.

\begin{equation}
\label{eq:fwhm}
\rm FWHM_{rest} = \sqrt{FWHM_{observed}^2 - FWHM_{instrument}^2}
\end{equation}

The observed Ly$\alpha$ profile has a near-symmetric shape with the main peak redshifted by only $\sim$60 km~s$^{-1}$ from the rest-frame central wavelength. The Ly$\alpha$ and He~II~$\lambda$1640 lines have rest-frame FWHM of $\sim$ 758$\pm$90 and 573$\pm$191 km~s$^{-1}$, respectively. The instrumental resolution of the LRIS spectrum is $\sim5\AA{}$ \citep{reddy2006}. We correct the observed line FWHM using equation \ref{eq:fwhm} and estimate the intrinsic values as 690 (Ly$\alpha$) and 526 (He~II~$\lambda$1640) km~s$^{-1}$. We measure the rest-frame EW of the He~II~$\lambda$1640 line as 8.3 $\AA$. The observed FWHM of the lines detected in the JWST grating spectra is nearly comparable to or smaller than the instrumental resolution at their respective wavelengths. The FWHM of He~I~$\lambda$10830 (i.e., 311$\pm$34 km~s$^{-1}$) is slightly larger than the instrumental resolution at the observed wavelength (i.e., $\sim$275 km~s$^{-1}$). The FWHM of the [SIII]~$\lambda$9069 line is also broader than the instrumental resolution; however, the associated error is significantly large, and hence we do not consider it to derive the intrinsic line width.

[OIII]~$\lambda$5007, the brightest among all the identified lines, is also detected in the Keck MOSFIRE spectrum (see Appendix \ref{s_add_spectra}, Figure \ref{fig:keck_mosdef} - bottom panel) which has a higher resolving power (instrumental resolution $\sim 5\AA{}$) than the NIRSpec. We derive the observed (intrinsic) FWHM of [OIII]~$\lambda$5007 as $\sim$169$\pm$10 (146) km~s$^{-1}$ from the Keck spectrum. We notice that the [OIII]~$\lambda$5007 line flux, derived from the Keck spectrum, is around two times higher than that estimated from the NIRSpec spectra. As we find no significant systematic offset in the continuum flux levels of the Keck $K$ band and NIRSpec G235M/F170LP spectra, which are corrected for slit-loss, this line flux difference could arise due the orientation of the MSA shutter, which covers approximately half of the galaxy G1, resulting in lower flux values. We further note here that the [OIII]~$\lambda$4959 line, in the Keck K-band spectrum, is significantly affected by noise, which resulted in a relatively lower FWHM than the [OIII]~$\lambda$5007 line (Table \ref{table_lines}). This produces a difference in the [OIII] line ratios derived from the Keck and the NIRSpec.

The H$\beta$ and H$\alpha$ Balmer lines, identified in the NIRSpec spectra, fall outside the wavelength coverage of the MOSFIRE, whereas the H$\gamma$ line is not identified in the Keck H-band spectrum. As a result, we could not constrain the intrinsic line width of the Balmer lines. The line flux and the rest-frame equivalent width of all the NIRSpec lines noted in Table \ref{table_lines} are robust, and the values could be even higher if the MSA covers G1 entirely inside. The [OII]~$\lambda$~3727/3729~\AA~ doublet, which is not covered by the NIRSpec gratings, is identified in both the NIRSpec prism (Figure \ref{fig:all_spectra}) and Keck MOSFIRE H band spectra (Figure \ref{fig:keck_mosdef} - top panel). We derive the O32 ratio (where O32 = $\frac{[OIII]~\lambda5007}{[OII]~\lambda\lambda3727,3729}$) using the line fluxes from the JWST spectra as the [OII]~$\lambda$~3727/3729~\AA~ line is marginally identified in the MOSFIRE spectra. We note here that the ratio of calibrated flux levels between the utilized NIRSpec prism and grating spectra is limited within $\lesssim$15\% (similar value reported by \citealt{deugenio2024} for spectra part of the JADES data release 3), which signifies negligible systematic bias in the derived O32 value. In Appendix \ref{s_flux_offset}, we provide a detailed discussion on the systematic error of different flux ratios derived in this study.

\begin{figure}
    \centering
    \includegraphics[width=3.6in]{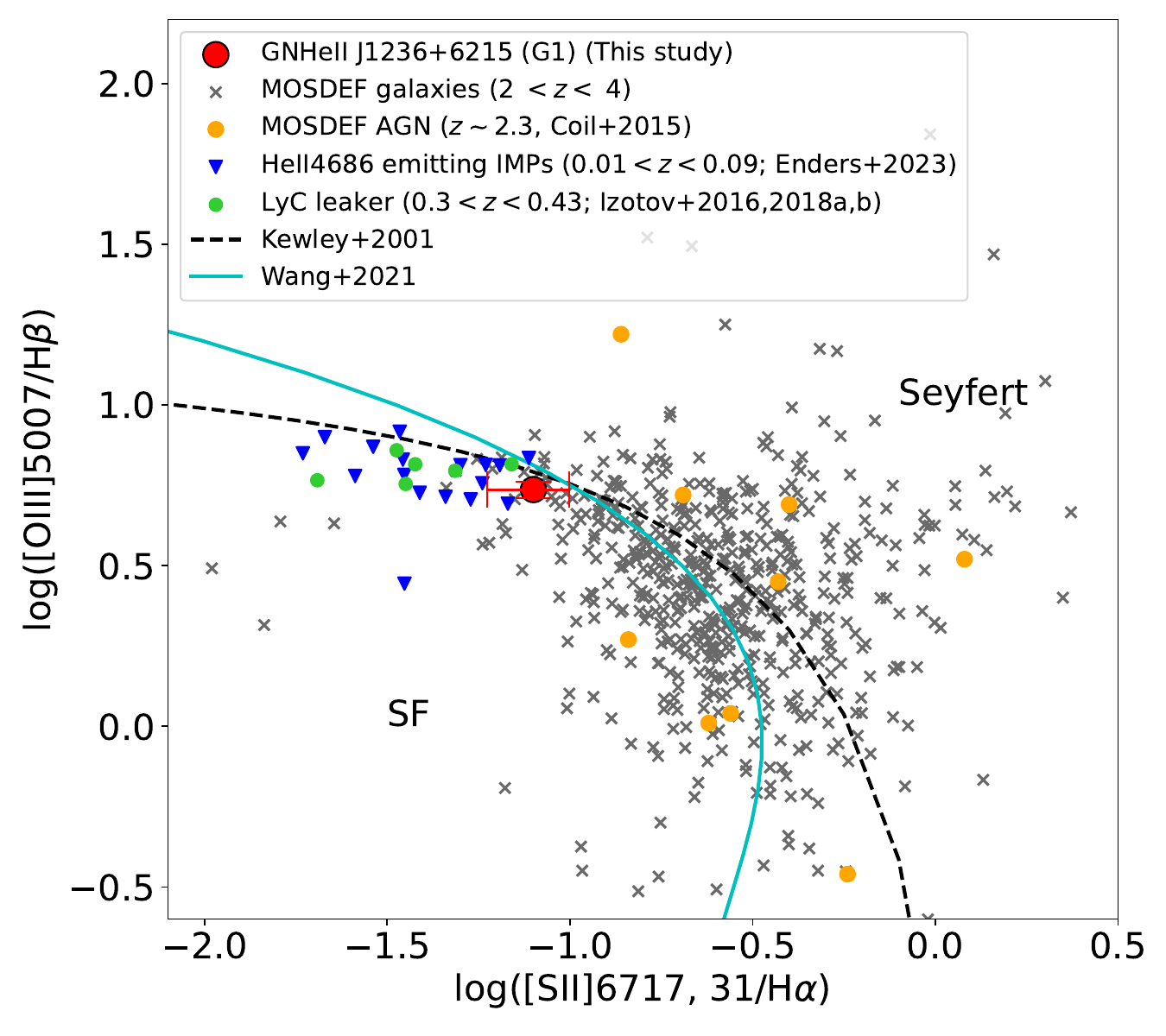} 
    \caption{[SII] BPT diagram that shows the location of galaxy GNHeII~J1236+6215 (red point) with respect to other populations. The grey crosses represent MOSDEF galaxy populations within redshifts 2 and 4 \citep{kriek2015}. The confirmed AGNs from the MOSDEF samples at $z\sim2.3$ \citep{coil2015} are shown in orange. The He~II~$\lambda$4686 emitting ionized metal-poor galaxies at $0.01< z <0.09$ from \citet{enders2023} are shown in blue. The green symbols denote the low redshift (0.29$< z <$0.43) LyC leakers reported in \citet{izotov2016,izotov2018,izotov2018b}. The black dashed line, taken from \citet{kewley2001}, shows the demarcation between Seyfert and star formation-type ionization. The cyan line denotes the locus of the star-forming galaxies used to define the ridge line for inferring [SII] deficiency by \citet{wang2021}.}
    \label{fig:bpt}
\end{figure}

\section{Discussion}
\label{s_discussion}

\subsection{Properties of GNHeII~J1236+6215}
\label{s_properties}
We use the photometric and spectroscopic measurements described in Section \ref{s_sed} and Section \ref{s_lines} to uncover the physical properties of the He~II emitting galaxy GNHeII~J1236+6215 (G1). We use the Galactic extinction-corrected HST F606W band flux (that samples the rest-frame FUV continuum) and derive the UV absolute magnitude (M$_{\rm UV}$) as $-22.09\pm0.02$ mag. The known He~II emitters between redshift $\sim$ 2 and 4, as reported by \citet{themia2019}, have M$_{\rm UV}$ ranging from $-18.85$ -- $-21.08$. This makes G1 one of the most UV luminous He~II emitters at $z\sim3$. Following the Balmer decrement method, we use the Galactic extinction-corrected H$\alpha$ and H$\beta$ line fluxes in equation \ref{eq:balmer} and derive interstellar reddening E(B$-$V) = 0.04$\pm$0.12, which suggests a lower dust extinction in the galaxy.

\begin{equation}
\label{eq:balmer}
    \rm E(B-V) = 1.97~log_{10}\left[\frac{(\frac{H\alpha}{H\beta})_{obs}}{2.86}\right]
\end{equation}
We use rest-frame FUV and different hydrogen recombination line luminosities to estimate galaxy's SFR utilizing scaling relations from \citet{murphy2011,zeimann2014,reddy2023}. The SFRs derived from the rest-frame FUV, H$\beta$, H$\alpha$, and Pa$\beta$ line fluxes, after correcting for both the Milky Way and galaxy's internal extinction, are 9.8$\pm$0.1, 7.6$\pm$0.4, 7.5$\pm$0.1, and 6.40$\pm$0.03 M$_{\odot}$~yr$^{-1}$ respectively, which agree well with the SED-derived value of 12.2$\pm$2.0 M$_{\odot}$~yr$^{-1}$ and confirm the star-forming nature of the galaxy G1. In Figure \ref{fig:sfr_mass}, we show the SFR and stellar mass of G1 along with the known He~II emitting galaxies and the main sequence relation of galaxies between redshift 2 and 4 from \citet{santini2017}. The galaxy G1 falls slightly above the main sequence populations at $z\sim3$ and within the distribution of known He~II emitters.

To understand the ionized state of the ISM, we utilize extinction-corrected [OIII]~$\lambda$5007, H$\beta$, [SII]~$\lambda$6718, 6732, and H$\alpha$ line fluxes and produce [SII] BPT diagnostic of the galaxy (Figure \ref{fig:bpt}). The measured line flux ratios, especially a low [SII]~$\lambda$6718, 6732/H$\alpha$ value, indicate ionization powered by star formation when compared with the diagnostic relation provided in \citet{kewley2001}. We also find that the line ratios reported for a sample of low-redshift LyC leakers \citep{izotov2016,izotov2018,izotov2018b} as well as He~II~$\lambda$4686 emitting ionized metal-poor (IMP) galaxies \citep{enders2023} closely overlap with that of G1 (Figure \ref{fig:bpt}). Besides, the location of G1 with respect to the locus of star-forming galaxies formulated by \citet{wang2021} as well as the distribution of MOSDEF galaxies and known AGNs \citep{kriek2015,coil2015}, as shown in Figure \ref{fig:bpt}, further supports the star-forming mode of ionization in G1. The value of S23, as defined in equation \ref{eq:s23}, is a well-known proxy to derive the nebular oxygen abundance in a galaxy. Using the empirical relation (equation \ref{eq:met}) between S23 and metallicity provided in \citet{montero2005}, we derive the gas-phase oxygen abundance of the galaxy as 12 + log(O/H) = 7.85$\pm$0.22, which translates to a metallicity Z $\sim$ 0.003 (i.e., $\sim$ 15\% of the Solar metallicity) signifying G1 to be a metal-poor galaxy. 
\begin{equation}
\label{eq:s23}
    \rm S23 = \frac{[SII]6718,6732 + [SIII]9069,9532}{H\beta}
\end{equation}

\begin{equation}
\label{eq:met}
    \rm 12 + log(O/H) = 8.15 + 1.85\times logS_{23} + 0.58\times (logS_{23})^2
\end{equation}
Utilizing another scaling relation between gas-phase metallicity and R23 (where, R23 = $ \rm \frac{[OIII]~\lambda5007 + [OIII]~\lambda4959 + [OII]~\lambda3727}{H\beta}$) provided in \citet{nakajima2022}, we derive a similar metallicity value of 12 + log(O/H) = 7.79$\pm$0.19 for the galaxy. To summarize, the He~II emitting galaxy G1 is a UV-luminous metal-poor star-forming galaxy with low dust content. All these properties are favorable to host massive young stars which can produce extreme ionizing photons to power He$^+$ ionization in the galaxy.

\subsection{Characteristics of the He~II~$\lambda$1640 line}
\label{s_properties_heii_line}

\begin{figure}
    \centering
    \includegraphics[width=3.5in]{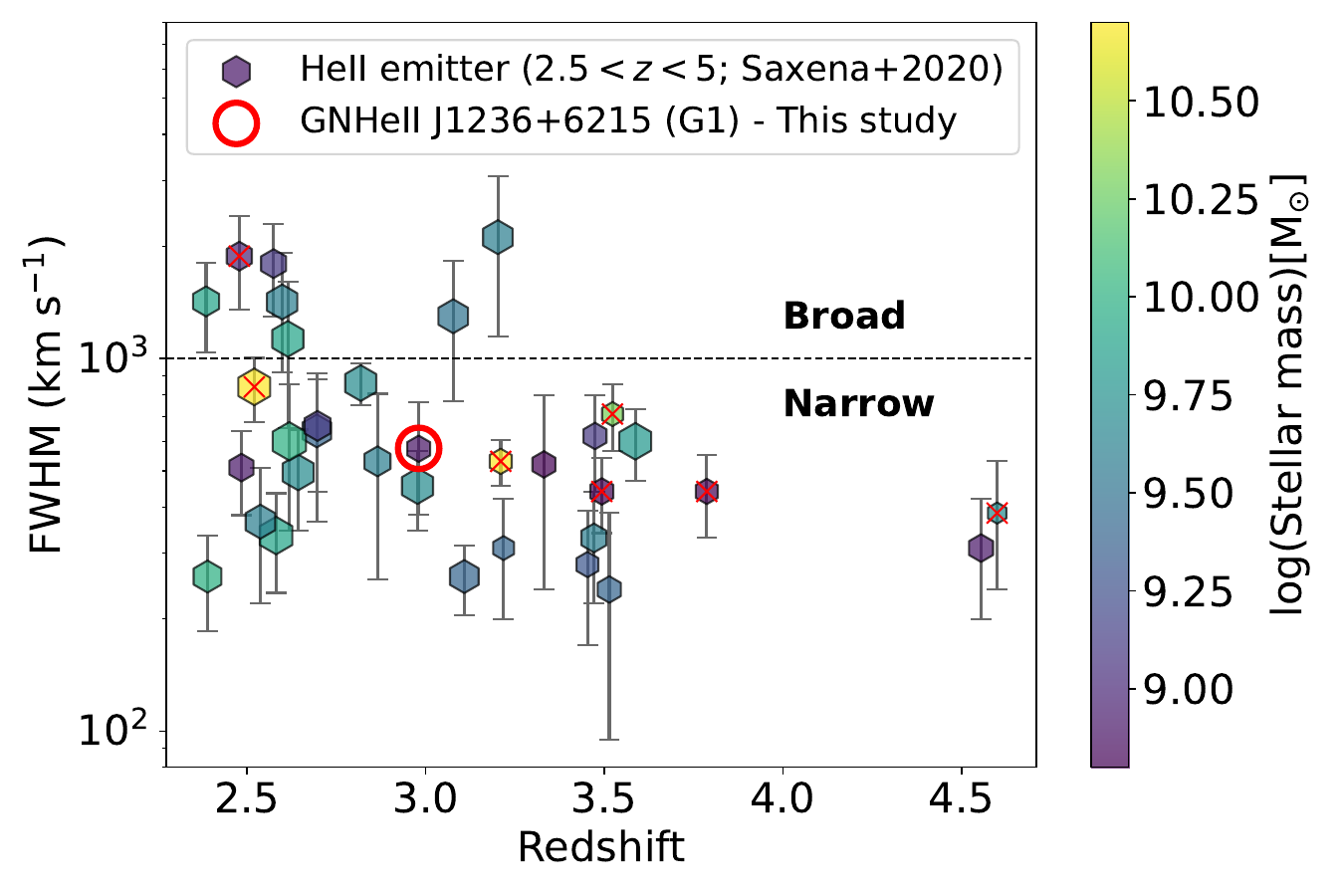}\\
\includegraphics[width=3.5in]{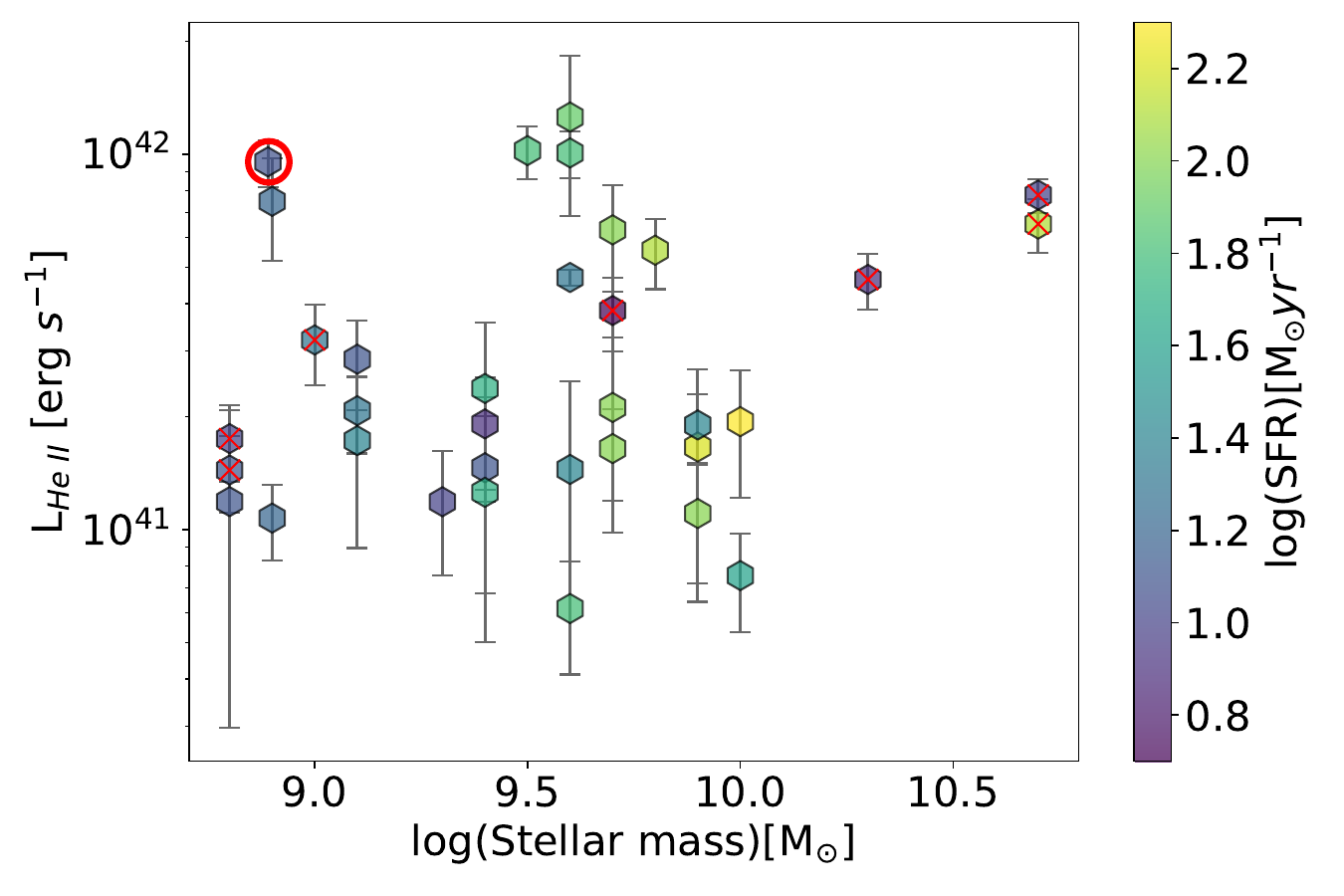}
    \caption{Properties of our identified He~II~$\lambda$1640 line is shown along with measurements of 33 He~II emitters (hexagonal markers) reported by \citet{saxena2020} between redshift $\sim$ 2.5 and 5. The galaxy GNHeII~J1236+6215 is marked with a red circle on both panels. \textbf{Top panel:} The distribution of redshift and FWHM of the detected He~II~$\lambda$1640 line. The colorbar represents the stellar mass of galaxies, while the size of the points is linearly scaled as per their log(SFR) between the range 0.7 and 2.1 M$_{\odot}$ yr$^{-1}$. The horizontal dashed line marks FWHM of 1000 km~s$^{-1}$ which is the boundary between narrow and broad He~II~$\lambda$1640 line adopted in \citet{saxena2020}. \textbf{Bottom panel:} The distribution of galaxy stellar mass and He~II~$\lambda$1640 line luminosity. The log(SFR)[M$_{\odot}$ yr$^{-1}$] is represented using the colorbar. The points marked with red crosses (in both panels) denote He~II-emitters classified as possible AGNs in \citet{saxena2020}. }
    \label{fig:literature}
\end{figure}

In Figure \ref{fig:literature}, we show our measurement for the He~II~$\lambda$1640 line of the galaxy GNHeII~J1236+6215 along with the sample of He~II emitters presented in \citet{saxena2020} from the VANDELS survey between redshifts 2 and 5. With reference to their classification of narrow (observed FWHM $<$ 1000 km~s$^{-1}$) and broad (observed FWHM $>$ 1000 km~s$^{-1}$) He~II emitters, the galaxy G1 falls under the former class with observed (deconvolved) FWHM of 573$\pm$191 (526) km~s$^{-1}$. \citet{cassata2013} also differentiated their He~II emitters between narrow and broad by considering the margin of observed FWHM as 1200 km~s$^{-1}$, which translates to a value of 663 km~s$^{-1}$ when corrected for VIMOS instrumental FWHM. They considered this value as a broad lower limit for He~II emission powered by WR stars. The intrinsic He~II~$\lambda$1640 line width of our galaxy (i.e., 526 km~s$^{-1}$) appears lower than this conservative limit which undermines the possible contribution of massive WR stars or AGN to produce it. \citet{saxena2020} reported that the stellar mass of both He~II emitters and non-emitters spans a similar range confirming no specific inclination over galaxy mass to power He$^+$ ionization. The galaxy G1 populates the lower side of the mass range covered by He~II emitters as reported by \citet{cassata2013,saxena2020} (Figure \ref{fig:sfr_mass}). The SFR of He~II emitters also covers a range similar to the star-forming non-emitters, with no specific trend of correlation found with other galaxy properties. The SFR of G1 falls on the lower side of the distribution reported by \citet{saxena2020}. The galaxy G1 appears to be one of the most luminous He~II emitters between redshift $\sim$ 2 and 5 with a He~II~$\lambda$1640 line luminosity of 9.55$\pm1.95\times 10^{41}$ erg s$^{-1}$ (Figure \ref{fig:literature}). This also agrees well with our estimated rest-frame EW of He~II~$\lambda$1640 line, which makes G1 one of the strongest He~II emitters at $z\sim3$ when comparing with the galaxies reported in \citet{themia2019,saxena2020}.

Apart from the He~II~$\lambda$1640 emission, we identify three more helium lines, i.e., He~I~$\lambda$5875, He~II~$\lambda$8236, and He~I~$\lambda$10830 in the NIRSpec spectra. The detection of He~II~$\lambda$8236 line, having the same ionization potential as He~II~$\lambda$1640, reinforces the presence of an extreme ionizing source in the galaxy. The He~I~$\lambda$10830 transition, brighter among the two, is known to have the strongest dependence on the electron density among the seven known He~I lines in the optical and near-infrared wavelength \citep{aver2015}. The detection of these additional helium lines makes G1 unique among the known high-redshift He~II emitters.

\subsection{Origin of He~II~$\lambda$1640 emission}
\label{s_heii_origin}

The He~II~$\lambda$1640 line profile itself carries important clues regarding the origin of extremely energetic photons (i.e., $\lambda < 228 \AA{}$) that could ionize He$^+$ to produce He~II~$\lambda$1640 transition. \citet{cassata2013} and \citet{saxena2020} utilize the He~II line properties of two different groups of He~II emitters between redshift $\sim$ 2 and 5 to understand the driver of He$^+$ ionization. The narrower FWHM of the He~II~$\lambda$1640 line in G1 (i.e., 573$\pm$191 km~s$^{-1}$) disfavours the possibility of an AGN which is expected to produce lines with a typical width of $\gtrsim1000$ km~s$^{-1}$ \citep{debreuck2000,cassata2013,saxena2020}. In addition to that, the detection of narrower Balmer lines (i.e., FWHM $\lesssim$300 km~s$^{-1}$) lowers the possibility of an AGN in G1. Apart from a broader line width, the presence of an AGN would also show high-ionization emission lines (particularly, C~IV~$\lambda$1549, N~V~$\lambda$1240; \citealt{shapley2003,themia2019}), which are not distinctly identified throughout the spectra of G1. Furthermore, the non-detection of G1 in the 2MS Chandra X-ray catalog \citep{alexander2003} in addition to the star-forming nature of ionization inferred from the [SII] BPT diagnostic further reduces the possibility of an AGN in the galaxy. Similarly, if the strong fast stellar winds of massive WR stars produce the He~II~$\lambda$1640 line emission, the expected line width would be broader (i.e., FWHM $\gtrsim$1500 km~s$^{-1}$; \citealt{shapley2003,shirazi2012}) than we identified. Moreover, WR stars-powered He~II emitter shows a P-Cygni CIV~$\lambda$1549 emission, which is even stronger than the He~II~$\lambda$1640 line in most cases. Therefore, the non-detection of high-ionization carbon or nitrogen lines and a narrower line width of the detected He~II~$\lambda$1640 and Balmer lines indicate that the He$^+$ ionization in the galaxy G1 is \textit{less likely due to either AGN or metal-rich WR stars}. 

The nebular He~II~$\lambda$1640 emission originating from the cooling of gravitationally infalling gas can produce a narrower He~II line that we observe in galaxy G1. The derived SFR and UV luminosity of G1 agree with this possibility, as a higher gas infall would also increase star formation in the galaxy. Detection of the Ly$\alpha$ line in G1 further supports the pristine gas-infalling case, as both Ly$\alpha$ and He~II~$\lambda$1640 transitions act as the major gas-coolant in the absence of heavier metals \citep{yang2006}. The other possibility to produce a narrower He~II~$\lambda$1640 line could be the Pop~III stars. However, the derived gas-phase metallicity of G1, although significantly low, contradicts the notable presence of metal-free Pop~III stars. Using hydrodynamical simulation, \citet{venditti2024} showed that due to the late accretion of pristine gas, Pop~III stars can still form in the outskirts of metal-enriched (i.e., 12 + log(O/H) reaching up to 7.7) galaxies with mass even higher than 3$\times$10$^9$ M$_{\odot}$ at redshift 6 - 7. \citet{sobral2015} reported such a scenario in a narrow-line He~II emitter identified at $z = 6.604$. Furthermore, considering inefficient mixing of metals, \citet{tornatore2007} and \citet{liu2020} found that the formation of Pop~III stars can continue even down to redshift 2.5 and 0, respectively. Considering the identified \HeII~emission, stellar mass, and gas-phase metallicity of the galaxy G1, this could indicate that a population of relatively metal-enriched stars contribute the identified Oxygen and Sulphur lines, whereas a smaller number of newly formed Pop~III stars can power He$^+$ ionization without majorly contributing to the observed stellar continuum. In this context, the non-detection of strong carbon and nitrogen lines indicates that the galaxy G1 could still hold pockets of pristine gas to form Pop~III stars. Assuming the observed \HeII\ line luminosity to be $\sim$1\% of the total PopIII bolometric luminosity (as considered in \citealt{wang2024,maiolino2024} also), we derive the total mass of the PopIII stellar populations in G1 as $\sim$7.7$\times$10$^5$ M$_{\odot}$ using the Eddington limit criteria. 

However, an alternative scenario proposed by \citet{grafener2015} showed that slow but strong wind of metal-poor (i.e., Z$<$0.1Z$_{\odot}$) VMS can also produce narrower He~II~$\lambda$1640 line (i.e., FWHM $\sim$300--500 km~s$^{-1}$) including weaker CIV~$\lambda$1549 emission. At relatively higher metallicity, VMS-powered \HeII\ emission of broader line-width (FWHM $\eqsim$ 1200 - 3200 km s$^{-1}$) is reported in 13 UV-bright star-forming galaxies at $z\sim2.2 - 3.6$ by \citet{upadhyaya2024}. \citet{schaerer2025} showed that inclusion of VMSs of mass up to 400 M$_{\odot}$ with a choice of flatter IMF can enhance the UV luminosity by 5-6 times from that of a normal SED produced using a Salpeter IMF of upper mass limit 100 M$_{\odot}$. The relatively higher UV luminosity (M$_{\rm UV}$ = $-22.09\pm0.02$) of GNHeII J1236+6215 could therefore indicate the plausible existence of VMSs in the galaxy, which have produced a \HeII\ line of high EW and also boosted the galaxy's rest-frame FUV flux, favoring a top-heavy IMF. Our observed UV continuum slope ($\beta_{\rm obs}$ = $-2.18\pm0.06$) also agrees well with models produced with VMSs of upper mass limit 300 - 400 M$_{\odot}$, including the nebular emission \citep{schaerer2025}. Considering the derived gas phase metallicity, it is therefore possible that the He$^{+}$ ionization in G1 is driven by such metal-poor VMS formed during the ongoing burst.

\begin{figure}
    \centering
    \includegraphics[width=3.5in]{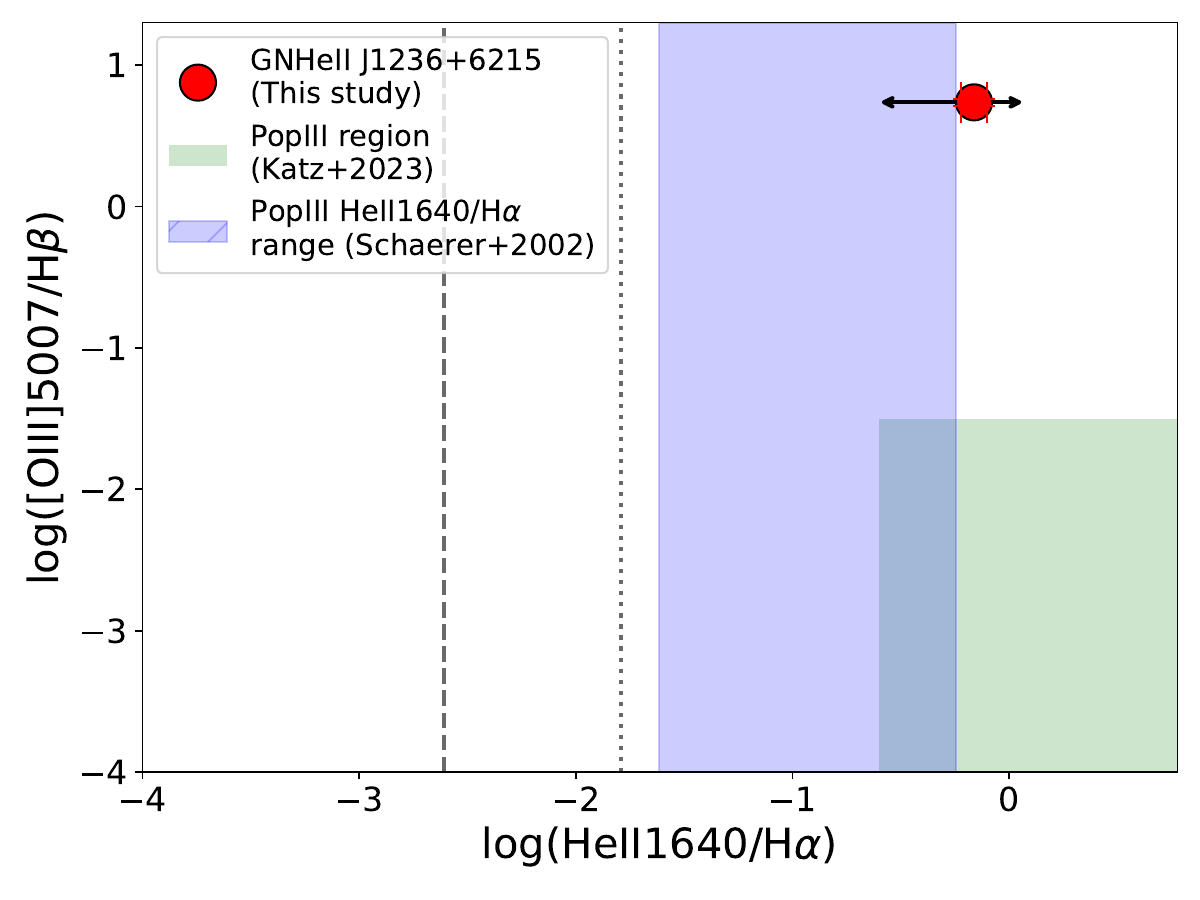}
    \caption{[OIII]~5007/H$\beta$ vs He~II~1640/H$\alpha$ line ratio of GNHeII J1236+6215 (red point). The black arrow indicates the uncertainty range in He~II~1640/H$\alpha$ arising from the systematic flux offset between the Keck and NIRSpec as discussed in Section \ref{s_heii_origin} and Appendix \ref{s_flux_offset}. The green patch highlights the range of [OIII]~5007/H$\beta$ and He~II~1640/H$\alpha$ values occupied by PopIII galaxies as simulated by \citet{katz2023}. The horizontal extent of the blue patch marks the range of He~II~1640/H$\alpha$ ratio produced for PopIII models of different stellar mass limits in \citet{schaerer2002}. The He~II~1640/H$\alpha$ value derived for solar metallicity (ZS: high mass loss) and 1/50 Z$_{\odot}$ low metallicity (ZL: mass loss) tracks from \citet{schaerer2002} are indicated using dotted and dashed gray lines, respectively.}
    \label{fig:popiii}
\end{figure}

As we find different drivers, including Pop~III stars, are capable of producing the observed \HeII~line in G1, we consider additional Pop~III diagnostic diagrams proposed in \citet{schaerer2002,katz2023,wang2024} for more insights. Utilizing cosmological radiation hydrodynamics simulations, \citet{katz2023} proposed a spectral hardness diagnostic diagram, constructed using [OIII]~5007/H$\beta$ and He~II1640/H$\alpha$ flux ratios, to distinguish ionization by Pop~III stars and more metal-enriched populations (Figure \ref{fig:popiii}). Similarly, \citet{wang2024} used MAPPINGS V photoionization models to distinguish between the contribution of normal metal-poor massive O-type stars and Pop~III stars in powering the He~II~$\lambda$1640 line they observed in a $z=8.16$ galaxy. Both their diagnostic grids could differentiate the contribution of Pop~III stars from the others in terms of higher He~II1640/H$\beta$ and He~II1640/H$\alpha$ values.

Using our derived spectral line fluxes, we estimate the O32, He~II1640/H$\beta$, He~II1640/H$\alpha$, and [OIII]~5007/H$\beta$ flux ratios as 7.28$\pm$0.11, 1.96$\pm$0.30, 0.69$\pm$0.10, and 5.45$\pm$0.32 respectively. The O32 and He~II1640/H$\beta$ of G1 closely match with the values reported for the He~II-emitting galaxy RX~J2129-z8He~II at $z=$8.16 by \citet{wang2024} and also agree well with the case of He$^+$ ionization by Pop~III stars. \citet{schaerer2002} reported such higher values of He~II1640/H$\beta$ and He~II1640/H$\alpha$ in the case of an instantaneous burst of Pop~III star formation with upper stellar mass limit extending up to 1000 M$_{\odot}$ added with strong mass loss (Figure \ref{fig:popiii}). The He~II1640/H$\alpha$ ratio of G1 also falls within the proposed range of ionization by Pop~III stars as reported in \citet{katz2023}, however the [OIII]~5007/H$\beta$ has a much higher value than what is expected from galaxies with only Pop~III stars (Figure \ref{fig:popiii}). This could indicate a similar scenario reported by \citet{sobral2015}, where the galaxy could host small pockets of Pop~III-like star formation along with relatively enriched normal populations with much larger mass fraction. The detection of multiple Oxygen and Sulphur emission lines supports this possibility for the galaxy G1. In that scenario, the derived flux ratios would represent the integrated contribution from both Pop~III and normal stellar populations. As ionization by normal massive stars would also efficiently produce hydrogen recombination lines, the actual He~II1640/H$\beta$ and He~II1640/H$\alpha$ flux ratios, in regions associated with only Pop~III stars, can therefore be even larger than the derived values. We note here that if we incorporate additional uncertainty into the extinction corrected line flux arising from the error in E(B$-$V), the error associated with the above flux ratios increases substantially to the values 1.96$\pm$1.30 and 0.69$\pm$0.43, respectively. Furthermore, as He~II1640 and both Balmer lines are observed using two different instruments, these ratios can have additional systematic uncertainty arising from the flux calibration bias as well as the contribution of G2 to each spectral continuum. However, we find that such systematic error creates only a smaller change to these flux ratios (see Appendix \ref{s_flux_offset} and Figure \ref{fig:spectra_offset}).

While pockets of Pop~III stars or metal-poor VMS can best explain the He$^+$ ionization in G1, the other source of hard ionizing photons to produce \HeII~line could also be the emission from X-ray binaries (XRBs) \citep{garnett1991,schaere2019}. The galaxy G1 is expected to host XRBs as their number increases with decreasing galaxy metallicity \citep{fornasini2019}. While we cannot completely rule out the contribution of XRBs to produce the observed \HeII~in G1, the non-detection of the galaxy in the 2MS Chandra X-ray catalog \citep{alexander2003} somewhat minimizes this possibility. Besides, the chance of supernova shock powering the observed \HeII~emission is again low, as we expect to see strong C~IV1549 emission also in that case \citep{allen2008}. To summarize, the He~II~$\lambda$1640 emission in G1 is most likely powered either by pockets of Pop~III stars or extremely metal-poor VMS. While XRBs or the gravitational cooling of infalling pristine gas could also produce such He~II~$\lambda$1640 line, the possibility of AGN or WR stars to drive He$^+$ ionization in G1 is rather low.

\subsection{A potential LyC leaker}
\label{s_lyc}
The detection of high ionization He~II, [OIII], [SIII] lines in G1 indicates that the galaxy must be producing enough ionizing photons which are also capable of ionizing neutral hydrogen atoms leading to an ISM transparent to LyC photons. In the rest frame of the galaxy G1, the HST F275W band falls entirely in the LyC range between wavelengths 550 and 800 $\AA{}$. Though we could not confirm the direct detection of ionizing LyC photons in the F275W band, we use the observed F275W flux (i.e., signal from escaping LyC photons) and the dust-corrected H$\alpha$ luminosity (which can be used to estimate amount of non-escaping LyC photons) to derive an upper limit of Lyman continuum escape fraction (f$_{esc}$) as 0.19 utilizing relations provided in \citet{saha2020}. We note here that the HST F275W filter has negligible red-leak (i.e., $<$1\%; \citealt{smith2018}) to create any significant bias on the derived f$_{esc}$ upper limit. However, the large photometric error associated with the F275W band flux of G1 makes this upper limit uncertain. 

Besides, we find G1 to host a favorable ISM condition that can allow LyC photons to escape the galaxy, making it a potential candidate for LyC leaker. Different studies have shown that galaxies showing Ly$\alpha$ emission, high O32 ratio, high SFR surface density, and having [SII] deficiency are prone to leak LyC photons \citep{anne2017,izotov2018,barrow2020,wang2021,kerutt2024}. In the case of galaxy G1, the higher value of O32 and [SIII]/[SII], the presence of hydrogen Balmer lines including H$\gamma$, and [SIII]9069,9532 lines indicates a higher ionized state of the ISM which supports the existence of extremely energetic stellar populations that can doubly ionize helium atom to produce He~II~$\lambda$1640 and He~II~$\lambda$8236 lines \citep{paswan2022}. The observed [SIII]/[SII] value infers a higher ionization potential, i.e., logU $\simeq -3$ - $-2$ when compared with the diagnostic diagrams of \citet{ramambason2020,kewley2019}. Besides, the non-detection of [OI]~$\lambda$6300 line further signifies that this high ionization in G1 is most likely not driven by radiative shock or AGN. Using the measured upper limit of [OI]~$\lambda$6300 line flux, we estimate an upper limit of [OI]/[OIII] flux ratio as 0.013, which is much smaller than the O32 value. This scenario supports density-bounded ionization by stellar source and disfavors contribution from shock or AGN \citep{plat2019}.

\begin{figure}
    \centering
    \includegraphics[width=3.5in]{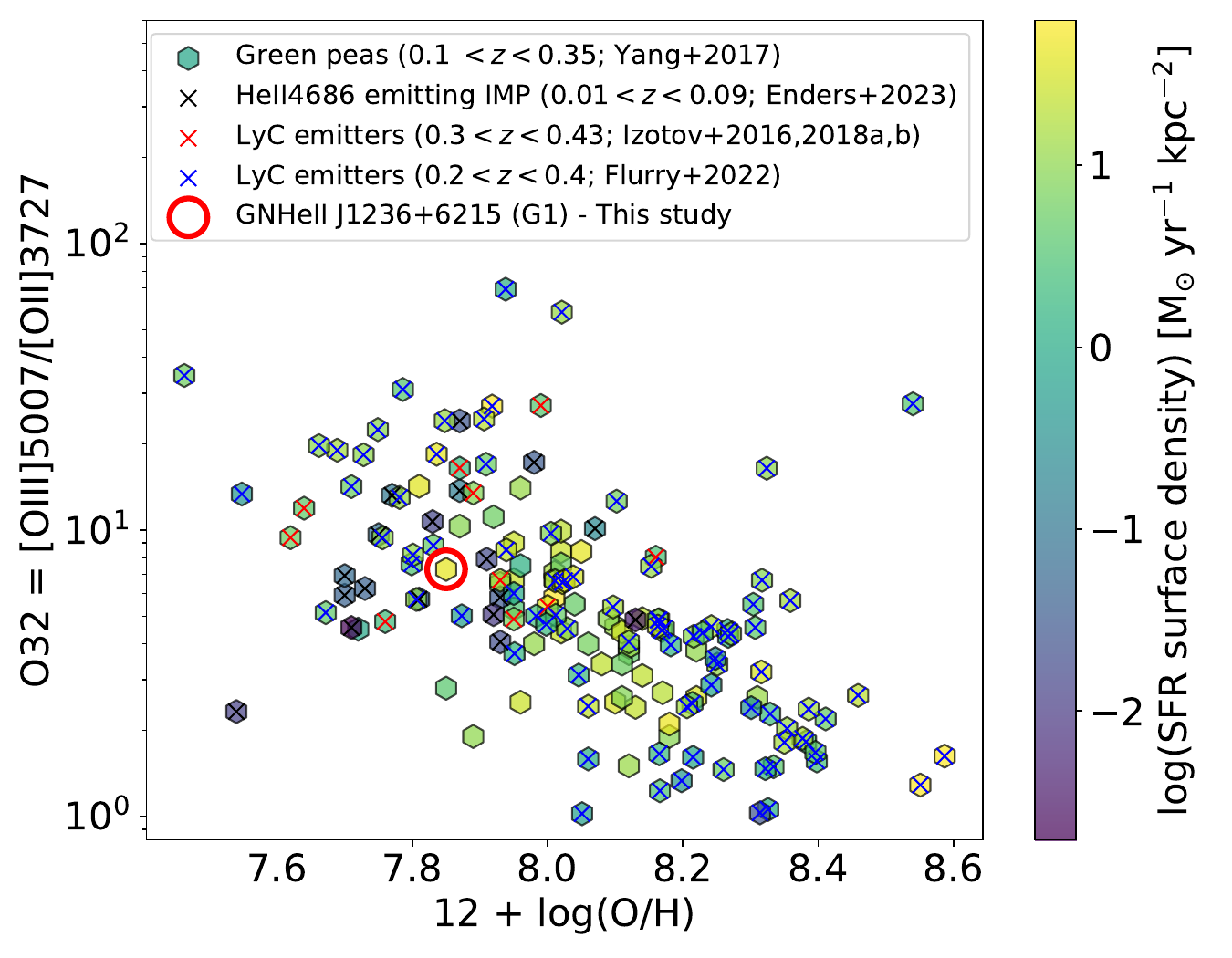}
    \caption{Gas-phase oxygen abundance 12 + log(O/H) and O32 ratio of the galaxy GNHeII~J1236+6215 (marked with red circle) along with a sample of green pea galaxies at $0.1 < z < 0.35$ from \citet{yang2017}, He~II~$\lambda$4686 emitting ionized metal-poor galaxies at $0.01< z <0.09$ from \citet{enders2023}, and low-redshift LyC leaking galaxies at 0.29$< z <$0.43 from \citet{izotov2016,izotov2018,izotov2018b} and at 0.2$< z <$0.4 from \citet{flury2022}. The SFR density of all the galaxies is highlighted using the color bar.}
    \label{fig:green_peas}
\end{figure}

In addition, the estimated values of O32, metallicity, E(B$-$V), SFR surface density, and stellar mass of G1 fall well within the regime of typical low redshift green pea \citep{yang2017} and ionised metal-poor \citep{enders2023} galaxies, which are known to host favorable ISM condition for leaking LyC photons (Figure \ref{fig:green_peas}). The observed properties of the known low-redshift LyC leakers \citep{izotov2016,izotov2018,izotov2018b,flury2022}, shown in Figure \ref{fig:green_peas}, also agree well with the derived metallicity, O32, and SFR surface density values of G1. Specifically, GNHeII~J1236+6215 shows a compact morphology that nearly matches the NIRCam PSF. Following the same approach adopted in \citet{fujimoto2025} for deriving size of such JWST-identified compact galaxies, we estimate an upper limit of effective radius as $\simeq$0.31 kpc, which infers a SFR surface density of $\sim$ 41.5 M$_{\odot}$ yr$^{-1}$ kpc$^{-2}$. Such compact nature and high SFR surface density further enhance the possibility of LyC leakage in G1 \citep{anne2017}. Besides, our best-fit SED infers an extremely blue UV continuum ($\beta_{0} = -2.71$) for the unattenuated stellar spectrum, indicating the presence of very young stellar populations that are efficient in producing ionizing photons \citep{zackrisson2013}. Additionally, both E(B$-$V) derived using Balmer decrement and the observed $\beta$ indicate a low dust extinction, which would further favor the escape of ionizing photons. In contrast to all these, the measured H$\beta$ EW of G1 (i.e., 26.6 $\AA{}$) falls towards the lower side of the distribution reported for low-redshift LyC emitters by \citet{flury2022}. However, considering the partial coverage of NIRSpec MSA on both G1 and G2, we restrict ourselves from interpreting LyC escape based on the H$\beta$ EW measurement.

Besides, the [SII]/H$\alpha$ line ratio of 0.08$\pm$0.02 places G1 in the regime of [SII]-deficient ionization ($\Delta$[SII] = $-$0.12, which is similar to the mean value of low-redshift LyC emitters presented in \citet{wang2021}), as formulated in \citealt{wang2019} (shown in Figure \ref{fig:bpt}), that favors the escape of energetic LyC photons from the galaxy \citep{wang2021,paswan2022}. As the ionization potential of [SII] line is smaller than that of the neutral hydrogen, a reduced [SII]/H$\alpha$ ratio signifies a density-bounded optically thin H~II region where the ionizing photons from massive stars can escape efficiently. Additionally, a near-symmetric Ly$\alpha$ profile with such a smaller shift ($\sim$60 km~s$^{-1}$) in the main peak plausibly indicates an optically thin ISM that favors the escape of LyC emission \citep{verhamme2015,anne2017}. Therefore, the ISM conditions of the galaxy GNHeII~J1236+6215, as inferred from multiple indicators, make it an ideal candidate for LyC leaker at $z\sim3$.

\section{Summary}
\label{s_summary}

We report the discovery of a low-mass metal-poor He~II~$\lambda$1640 emitting galaxy GNHeII~J1236+6215 at $z = 2.9803$ in the GOODS-north field. Considering the limited number of confirmed He~II~$\lambda$1640 emitters beyond redshift 2, this study adds an important object to the known sample. Utilizing the rich photometric and spectroscopic data from the Keck, JWST, and HST, we find that the galaxy has lower stellar mass (M = 7.8$\pm3.1\times$10$^8$ M$_{\odot}$), minimal dust content (E(B$-$V) = 0.04$\pm$0.12 \& observed $\beta = -2.18\pm0.06$), lower gas-phase metallicity (12 + log(O/H) = 7.85$\pm$0.22), and a higher star formation rate (SFR$_{\rm SED}$ = 12.2$\pm$2.0 M$_{\odot}$~yr$^{-1}$) including a highly ionized ISM (O32 = 7.28$\pm$0.11 and [SIII]/[SII] = 1.97$\pm$0.48). The galaxy appears as one of the most luminous He~II~$\lambda$1640 emitter (L$_{\rm He~II} = 9.55\pm1.95\times10^{41}$ erg s$^{-1}$) with a relatively narrower line width (i.e., FWHM $\sim$ 573$\pm$191 km~s$^{-1}$) at $z \sim 2 - 5$.

Our results show that the ionization by Pop~III stars formed in small pockets of pristine gas or metal-poor VMSs formed during the ongoing burst could best explain the narrower \HeII~line-width in G1. The observed O32, He~II1640/H$\beta$, and He~II1640/H$\alpha$ line ratios of G1 agree well with the case of Pop~III stars powering the He~II line as inferred from diagnostic diagrams provided in \citet{schaerer2002,katz2023,wang2024}. However, the relatively higher value of [OIII]~$\lambda$5007/H$\beta$ and the gas-phase metallicity disfavor pristine PopIII-like conditions in the galaxy. Considering the possibility of Pop~III star formation at lower redshift proposed in \citet{tornatore2007,liu2020} and the previous discovery of a galaxy hosting Pop~III-like stars alongside normal stellar population by \citet{sobral2015}, we speculate that the galaxy G1 could plausibly host small pockets of Pop~III stars, which contribute $\lambda<228\AA{}$ energetic photons to produce the observed He~II~$\lambda$1640 emission, whereas the bulk of its stellar populations are relatively metal enriched and they represent the derived metallicity and other physical parameters of G1. The narrower \HeII\ line width, low metallicity, and the high UV luminosity also favor metal-poor VMSs as the plausible driver of He$^{+}$ ionization in GNHeII~J1236+6215. Besides, the cooling of pristine infalling gas also fits well with the observed \HeII~and Ly$\alpha$ emission. On the other hand, the detection of narrower He~II~$\lambda$1640 and Balmer lines in addition to the non-detection of high-ionization carbon lines (eg, C~IV~$\lambda$1549) and X-ray emission indicates that the He$^+$ ionization in the galaxy G1 is less likely due to AGN or massive WR stars. While we could not rule out the possible contribution of XRBs to power the \HeII~emission, the non-detection of X-ray emission from G1 undermines this possibility.

A significant [SII]-deficiency (with [SII]/H$\alpha \sim$0.08$\pm$0.02 and $\Delta$[SII] = $-$0.12) could further indicate the presence of density-bounded optically thin H~II regions, which favor the escape of Lyman continuum photons from the galaxy G1. This further agrees with the compact morphology (r$_e<0.31$ kpc), lower E(B$-$V), higher O32, [SIII]/[SII], SFR surface density values, and the observed Ly$\alpha$ line profile with a smaller shift in the main peak from the systemic redshift, indicating a plausible scenario of LyC escape from optically thin and dust-poor ISM. Therefore, the He~II emitter reported in this study not only provides insights into different driving mechanisms to produce this high-ionization line, but it also highlights a favorable ISM condition for the leakage of ionizing LyC photons from the galaxy. This could also indicate a plausible connection between the production of energetic ionizing photons and their escape from the ISM, making He~II~$\lambda$1640 emitters important in the context of studying LyC leakers to understand reionization. Finally, our study also demonstrates the strength of JWST NIRCam imaging over the best available HST images for dealing with the foreground interloper problem while studying high-redshift galaxies in deep fields.

\section*{Acknowledgements}
This work is supported in part by the National Science and Technology Council of Taiwan (grant NSTC 112-2112-M-001-027-MY3) and the Academia Sinica Investigator Award (grant AS-IA112-M04). C.-C.C. acknowledges support from the National Science and Technology Council of Taiwan (111-2112-M-001-045-MY3), as well as Academia Sinica through the Career Development Award (AS-CDA-112-M02). Finally, we thank the anonymous referee for valuable suggestions.

\software{SAOImageDS9 \citep{joye2003}, Matplotlib \citep{matplotlib2007}, Astropy \citep{astropy2013,astropy2018}, EAZY \citep{brammer2008}, T-PHOT \citep{merlin2015,merlin2016}, CIGALE \citep{boquien2019}}


\appendix

\counterwithin{table}{section}
\counterwithin{figure}{section}

\section{Photometry method}
\label{s_tphot}

To match the image resolution, we consider the JWST F444W band, having the broadest PSF (i.e., FWHM $\sim$ 0\farcs145), as the reference and produce matching kernels for all other bands utilizing PSF images in each corresponding band. For each different HST and JWST band, we apply a \textsc{CosineBell} window with $\alpha$ values as adopted in \citet{rieke2023} while producing respective matching kernels. Finally, we convolve each observed image with the corresponding matching kernel using \textsc{convolve\_fft} Astropy package to produce PSF-matched images in all the bands. 

Once we have matched the multiband images to the resolution of the JWST F444W band, to ensure consistent colors, we find that there was sufficient blending between G1 and G2 such that they could not be separated using Source Extractor \citep{bertin_arnouts1996}. On the other hand, performing forced photometry on such low-resolution images has the disadvantage of contamination. Therefore, to ensure photometric accuracy and to remove the effects of the PSF, we first tried the 2D image modeling using \textsc{GALFIT} \citep{peng2002}. We model G1 with Sersic/PSF components and G2 with an exponential disk, but in most cases, the fitting was either not good or yielded unphysical structural parameters. 

We then use T-PHOT, v2.0 \citep{merlin2015,merlin2016}, which can perform unbiased multi-band photometry over a wide range of spatial resolutions. It uses prior information such as the position, shape, and size of objects in a region of interest from a high-resolution image and performs photometric measurements on the corresponding low-resolution images. We prepare a mean of the JWST SW filter images and use that as the high-resolution prior in our case. We then generate a kernel using the corresponding mean, high-resolution PSF, and the F444W PSF which will degrade the prior image to the resolution of the latter. After this, we obtain a segmentation image and catalog of sources, from the prior image, in our region of interest using Source Extractor. All the above information is then fed into a parameter file to be used by T-PHOT to perform PSF-matched photometry. The basic objective of T-PHOT is to obtain the fluxes of the low-resolution counterparts of sources detected in the high-resolution prior. It performs minimization to obtain a scale factor that will accordingly adjust the amplitude of the prior light distribution in the high-resolution image to match that in the PSF-matched image. We then compute the fluxes of G1 and G2, from the PSF-matched images, (listed in Table \ref{table_photometry}) to be used for the SED modeling.

\section{Additional spectra}
\label{s_add_spectra}
\begin{figure*}[h!]
    \centering
    \includegraphics[width=6.5in]{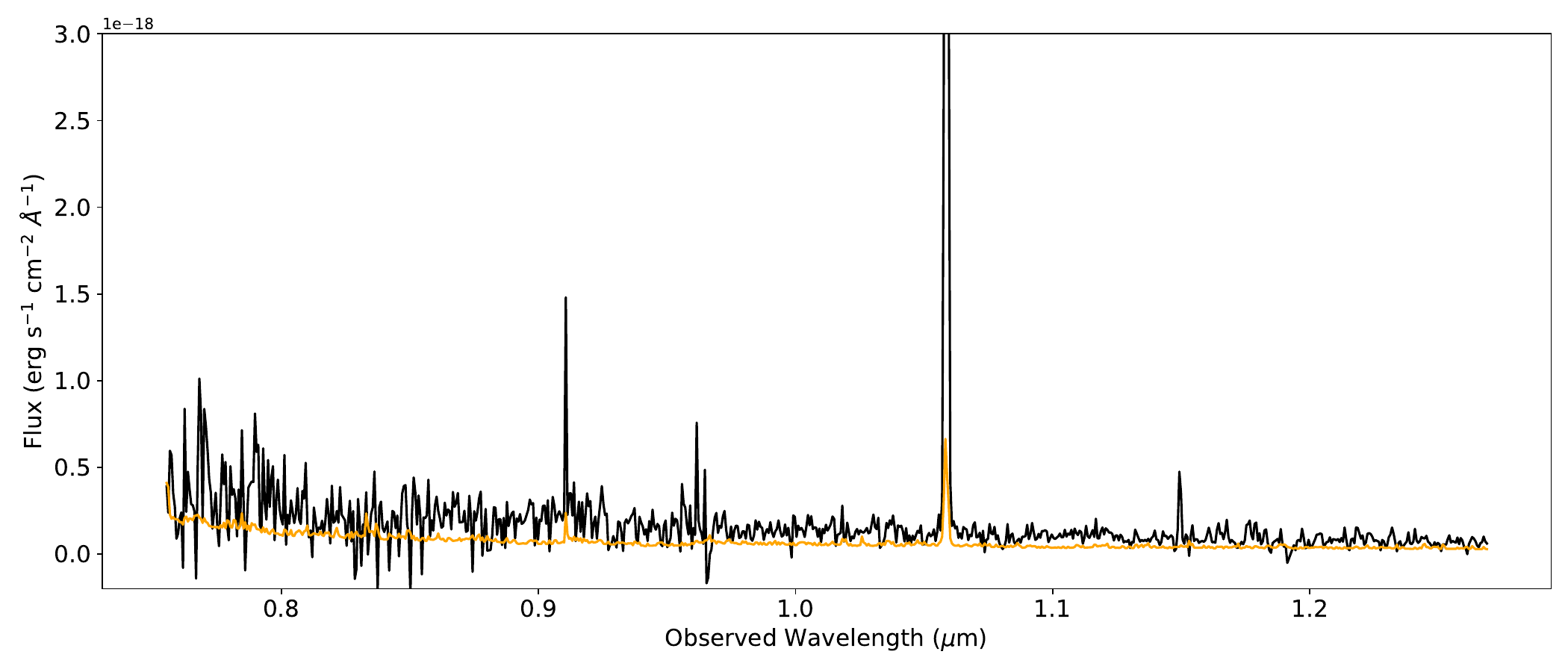} 
    \caption{Observed JWST NIRSpec/MSA spectra (black) of the galaxy including the error in flux values (orange) acquired with the grating G140M/F070LP.}
    \label{fig:jwst_f070lp}
\end{figure*}

\begin{figure*}
    \centering
    \includegraphics[width=6.5in]{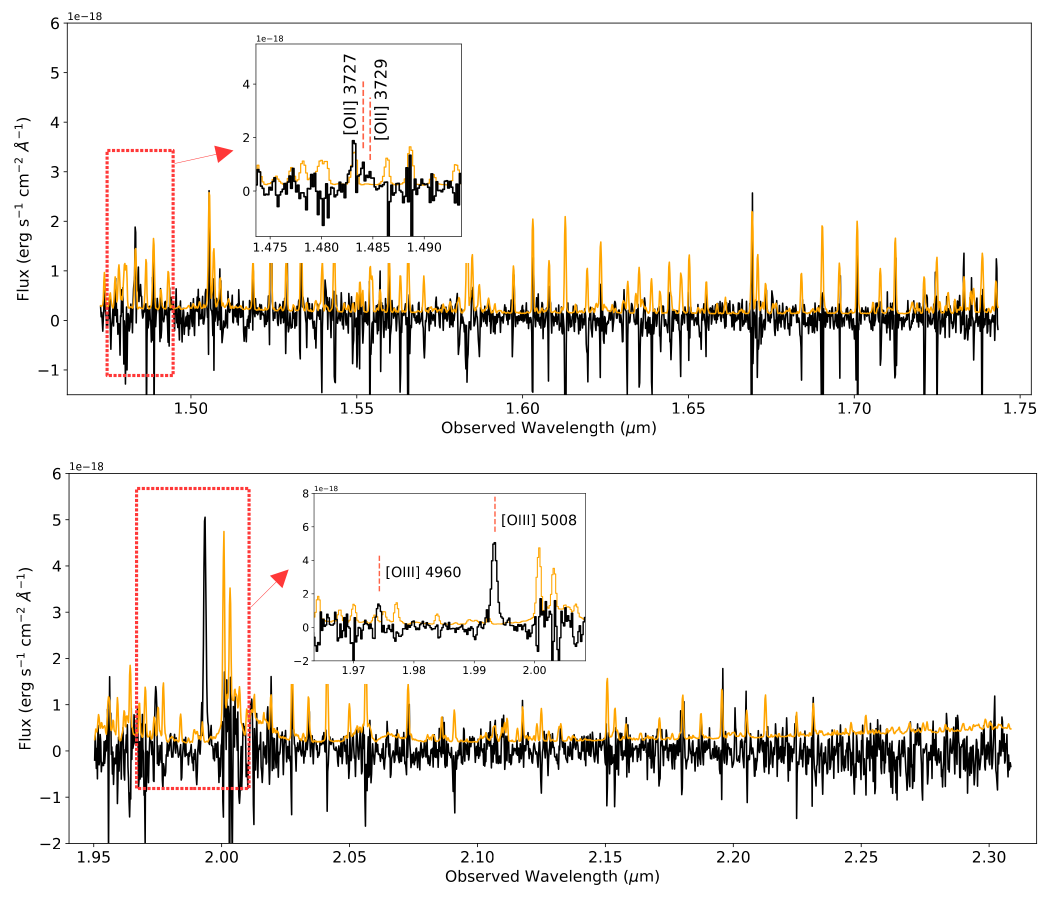} 
    \caption{Observed Keck MOSFIRE \textit{H} (top panel) and \textit{K$_s$} (bottom panel) band spectra of the galaxy (black) including the corresponding error in flux values (orange). The [OII]~$\lambda$3727/3729 lines, shown in the inset of the top panel, are marginally detected in the H band spectra. The K band spectrum shows clean detection of the [OIII]~$\lambda$4959 and [OIII]~$\lambda$5007 lines, which are also identified in the JWST G235M/F170LP grating spectrum (Figure \ref{fig:all_spectra}).}
    \label{fig:keck_mosdef}
\end{figure*}

In Figure \ref{fig:jwst_f070lp}, we show the observed NIRSpec G140M/F070LP spectrum. Considering the redshift of the galaxy G1, we do not identify any emission line. The two brightest line features, observed in the spectra, have larger errors in the associated flux values, which indicates those to be potential artifacts. The Keck MOSFIRE H and K band spectra are shown in Figure \ref{fig:keck_mosdef}. We identify three lines, i.e., [OII]~$\lambda$3727/3729, [OIII]~$\lambda$4959, and [OIII]~$\lambda$5007 in the MOSFIRE spectra. The line properties are included in Table \ref{table_lines}.

\section{JWST NIRSpec 2D spectra}
\label{s_jwst_2d}

\begin{figure*}[h!]
    \centering
    \includegraphics[width=6.0in]{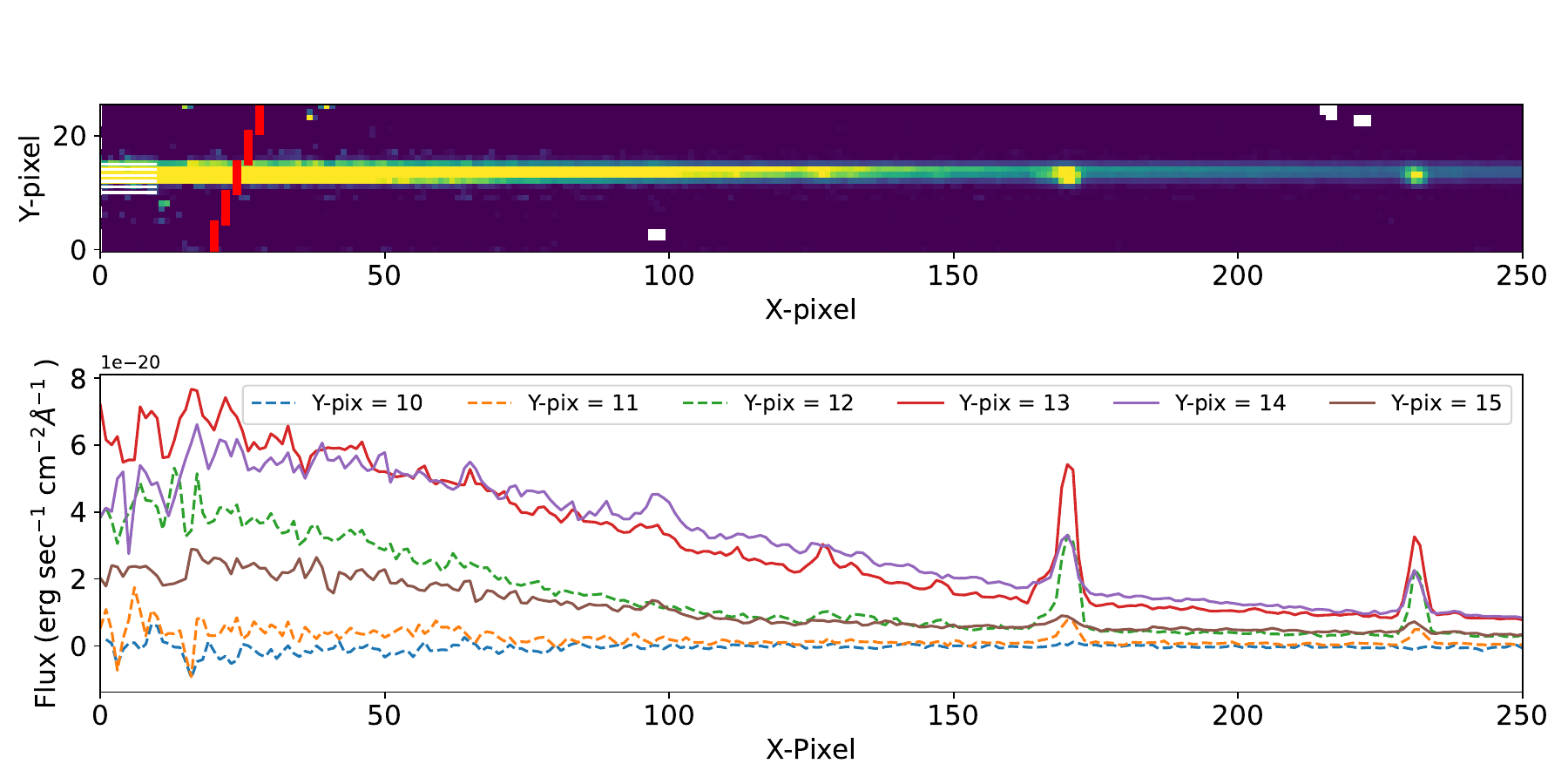} 
    \caption{The bluer half of the NIRSpec prism 2D spectrum is shown on the upper panel. The extent of five shutters along the spatial axis is shown by five red lines arbitrarily placed on the 2D spectra for better visualization. We extract spectra along six different pixel rows (Y-pixel = 10,11,12,13,14,15) which are marked in white horizontal lines on the left part of the 2D spectrum. The extracted spectra, shown in the bottom panel, highlight the relative strength of the continuum and emission lines along each pixel row.}
    \label{fig:jwst_prism_2d}
\end{figure*}

\begin{figure*}[h!]
    \centering
    \includegraphics[width=6.0in]{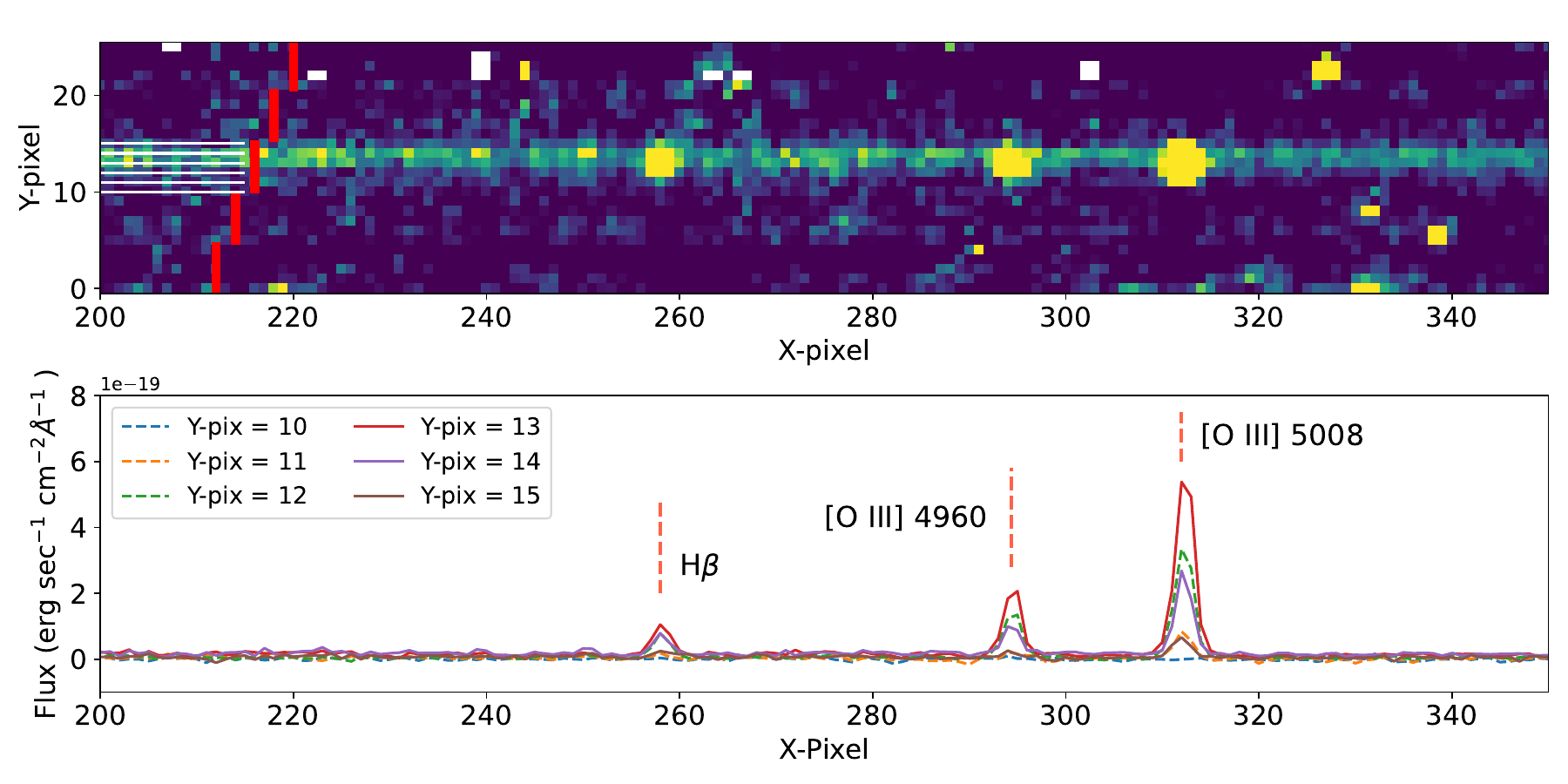} 
    \caption{A selected part of the NIRSpec G235M/F170LP grating 2D spectrum that contains three of the identified emission lines (i.e., H$\beta$, [OIII]~4959, 5007 is shown on the upper panel. The extent of five shutters along the spatial axis is shown by five red lines arbitrarily placed on the 2D spectra for better visualization. We extract spectra along six different pixel rows (Y-pixel = 10,11,12,13,14,15), which are marked in white horizontal lines on the left part of the 2D spectrum. The extracted spectra are shown in the bottom panel, where we also mark three emission lines to highlight their relative strength along each pixel row.}
    \label{fig:jwst_f170_2d}
\end{figure*}

We utilize the JWST NIRSpec 2D spectra to further investigate spectral profiles spatially along the MSA length. Considering the orientation of the MSA (Figure \ref{fig:color_image}) as provided in the header as well as in the corresponding APT file, galaxy G2 spans from the middle to the lower half of the central shutter, whereas galaxy G1, which is almost half-covered, sits mostly on the upper half. We use the 2D spectrum taken with NIRSpec prism and G235M/F170LP grating to investigate it further. In the upper panel of Figure \ref{fig:jwst_prism_2d}, we have shown the bluer half of the NIRSpec prism 2D spectrum. The spatial extent of five shutters across the dispersion axis is also marked. The continuum of the dispersed light is noticed to be effectively distributed from the middle to the upper half of the central shutter. The central MSA shutter spans within 10 - 16 pixels on the image plane. We extract spectra for six different pixel rows (Y-pix = 10 - 15, marked by white horizontal lines on the 2D spectra) and show them in the bottom panel of the same figure. The continuum becomes brightest and steepest at the 13th-pixel row and falls more rapidly towards the lower half than the upper half. This indicates that the galaxy G1 has a larger contribution to the spectral continuum of the archival 1D spectra analysed in this study than G2. We use a part of the NIRSpec G235M/F170LP grating spectrum that covers three of the detected emission lines (H$\beta$, [OIII]~$\lambda$4959, and [OIII]~$\lambda$5007) to examine further (Figure \ref{fig:jwst_f170_2d}). The 2D morphology of the emission lines agrees well with the point-like appearance of the galaxy G1. All three emission lines become brightest at the 13th-pixel row following a gradual decrease on either side. The orientation of the MSA shutters, as shown in vertical red lines, shows that the emission line flux distribution is slightly more inclined to the upper half of the central shutter, confirming G1 as the associated source. However, we cannot entirely rule out the contribution from the lower half of the shutter. This would indicate that the emission from a part of the lower half is either coming from the flux distributed in G1's wing (as normally observed in JWST PSF) or that the galaxy G2 is also located at the same redshift as G1 and also contributes to the observed emission lines. The latter possibility would also mean that galaxy G2 is a confirmed LyC leaker located at $z=2.9803$.

\section{Offset in photometric and spectral flux levels}
\label{s_flux_offset}

\begin{figure*}[h!]
    \centering
    \includegraphics[width=7.0in]{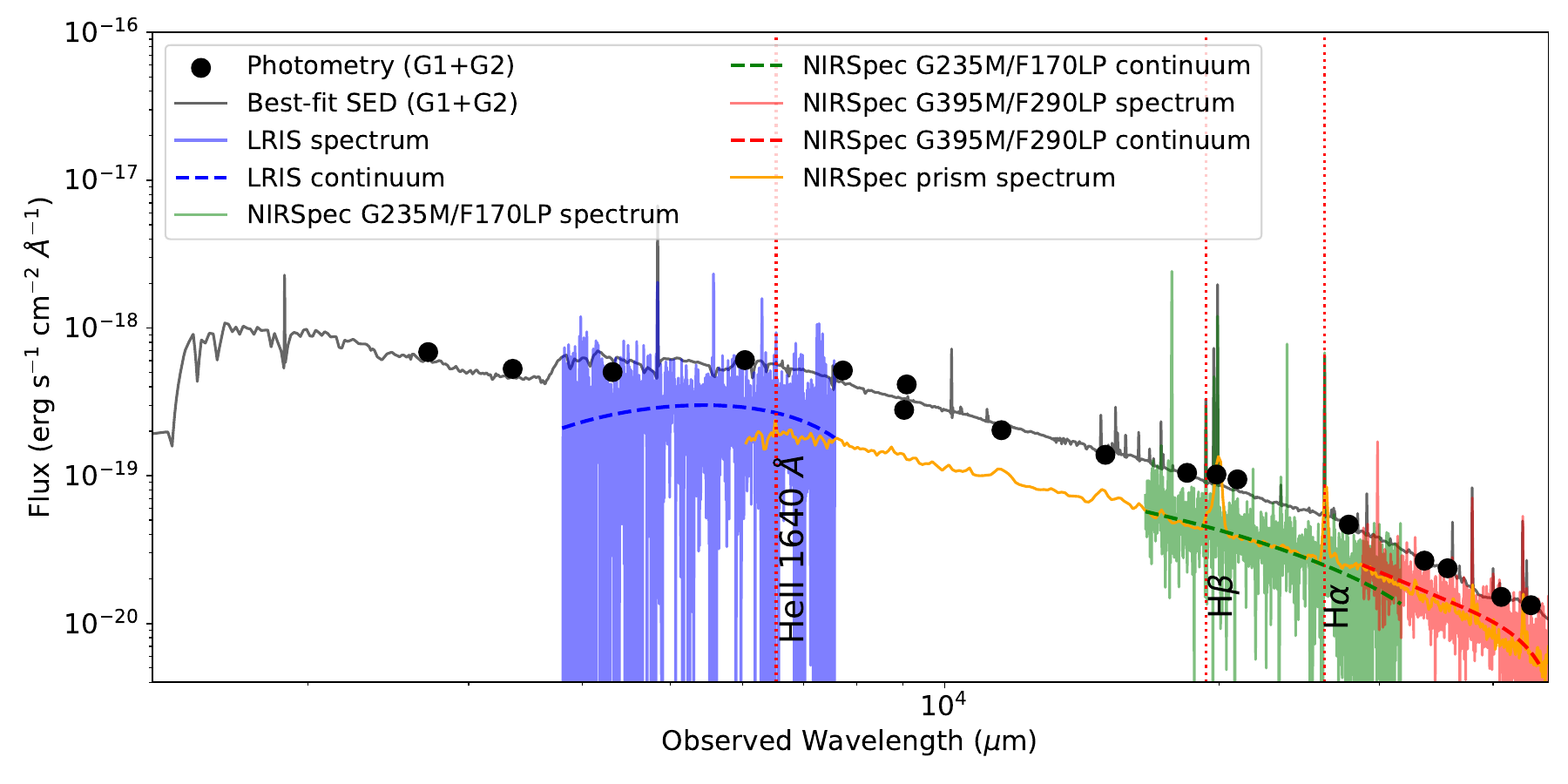} 
    \caption{The figure shows the relative difference in the flux levels of different spectra with respect to the combined photometric flux levels of the galaxies. The summed observed fluxes of G1 and G2 in all the filters are shown in black points. The grey SED represents flux values combining the best-fit SEDs of G1 and G2 in the observed frame. The Keck LRIS, NIRSpec G235M/F170LP, and NIRSpec G395M/F290LP spectra (from which all the emission lines are identified) are shown in blue, green, and red, respectively. The fitted continuum flux levels for each spectrum are displayed using a dashed line of the same color. The vertical red dotted lines mark the location of three emission lines: He~II1640, H$\beta$, and H$\alpha$. The flux levels of all three spectral continuum are lower than the combined photometric flux levels by a similar factor between $\sim$2.07 and 2.31. The NIRSpec prism spectrum is shown in orange to highlight the difference in flux levels between the NIRSpec gratings and the prism.}
    \label{fig:spectra_offset}
\end{figure*}

We utilize different emission line flux ratios for various scientific interpretations in this study. As the spectra used in our analysis are acquired with different instruments, any systematic offset in their flux levels could affect the related derived numbers. Among all flux ratios, He~II1640/H$\beta$ \& He~II1640/H$\alpha$, which are utilized for the interpretation of Pop~III stars, could have the maximum systematic bias as each of these ratios involves emission lines observed with two different telescopes (i.e., Keck LRIS for He~II1640 and JWST NIRSpec for H$\beta$ and H$\alpha$). The difference in the flux levels of each spectrum from their actual expected value may originate from two factors: (i) relative difference arising from the flux calibration of each spectrum, (ii) contribution of the galaxy G2 to the observed continuum of each respective 1D spectrum. To quantify the combined effect of both factors on these two line ratios, we estimate the difference between the fitted spectral continuum (i.e., LRIS spectrum for \HeII~and NIRSpec G235M/F170LP spectrum for H$\beta$ and H$\alpha$) and the photometric continuum derived from the combined best-fit SEDs of galaxies G1 and G2 at the observed wavelength of these three lines (demonstrated in Figure \ref{fig:spectra_offset}). We find that the combined photometric flux level is higher than the spectral continuum level by a factor of $\sim$2.31$\pm$1.03, 2.07$\pm$0.92, and 2.18$\pm$1.09 at the wavelength of He~II1640, H$\beta$, and H$\alpha$, respectively. The associated errors are derived from the 1$\sigma$ error in the spectral continuum flux values at each respective wavelength. These estimated offset factors indicate that the value of He~II1640/H$\beta$ and He~II1640/H$\alpha$ could change only by a factor of 1.15$\pm$0.70 and 1.06$\pm$0.71 if we rescale the spectral flux levels to the observed combined photometric flux level of the galaxies.

The flux ratios used in the [SII] BPT diagnostic (Figure \ref{fig:bpt}) are not affected by such systematic bias as all the emission lines are sampled from one NIRSpec grating. We find the flux levels between NIRSpec prism and two gratings (G235M/F170LP and G395M/F290LP) 1D spectra differ by $\lesssim$15\% (Figure \ref{fig:spectra_offset}; similar value is also reported by \citet{deugenio2024} in the JADES data release paper), which signifies minor change in O32, S23, and [SIII]/[SII] values due to this systematic offset.

\end{document}